\documentclass[10pt]{article}
\usepackage[colorlinks,
            linkcolor=blue,
            urlcolor=blue,
            anchorcolor=red,
            citecolor=blue
            ]{hyperref}
\usepackage{graphicx}
\usepackage{cite}
\usepackage{subfigure}
\usepackage{epstopdf}
\usepackage{rotating}
\usepackage{amsmath}
\usepackage{amssymb}
\usepackage{float}
\usepackage{color, soul}

\begin{document}
Submitted to {\it{Chinese Physics C}}
\begin{center}
{\Large \bf Centrality-dependent chemical potentials of light hadrons and quarks based on transverse momentum spectra and particle yield ratios in Au-Au collisions}

\vskip1.0cm

Xing-Wei He$^{a}$, Hua-Rong Wei$^{a,}${\footnote{E-mail: huarongwei@qq.com;huarongwei@lsu.edu.cn}},
Bi-Hai Hong$^{a,}${\footnote{E-mail: hbh{\_}201409@126.com}}, Hong-Yu Wu$^{a}$, Wei-Ting Zhu$^{a}$, Feng-Min Wu$^{b,c}$, and  Fu-Hu Liu$^{c}$  

{\small\it $^a$Institute of Optoelectronic Technology, Lishui University, Lishui, Zhejiang 323000, People's Republic of China\\
$^{b}$Department of Physics, Zhejiang Sci-Tech University, Hangzhou, Zhejiang 310000, People's Republic of China\\
$^c$Institute of Theoretical Physics \& State Key Laboratory of Quantum Optics and Quantum Optics Devices, Shanxi University, Taiyuan, Shanxi 030006, People's Republic of China}
\end{center}

\vskip1.0cm

{\bf Abstract:} We describe the transverse momentum spectra of $\pi^\pm$, $K^\pm$, $p$, and $\bar{p}$ produced in different centralities gold-gold (Au-Au) collisions at different collision energies range from 7.7 to 62.4 GeV by using a two-component Erlang distribution in the framework of multi-source thermal model. The fitting results are consistent with the experimental data, and the centrality- and energy-dependent yield ratios of negative to positive particles are obtained from the normalization constants. Based on the yield ratios, the energy- and centrality-dependent chemical potentials of light hadrons ($\pi$, $K$, and $p$) and quarks ($u$, $d$, and $s$) are extracted. The study shows that the dependences of the three types of particle yield ratios on centrality are not significant, especially for $\pi$. While the logarithms of the three yield ratios show obvious linear dependence on $1/\sqrt{s_{NN}}$ over a range from 7.7 to 62.4 GeV. The extracted chemical potentials (the absolute magnitude for $\pi$) of light hadrons and quarks show obvious dependence on energy, and decrease with the increase of energy over a range from 7.7 to 62.4 GeV. The dependences of the energy-dependent chemical potentials of light hadrons and quarks on centrality are relatively more obvious in low energy region than that in high energy region. The derived curves of chemical potentials for all centralities, from the linear fits of the logarithms of yield ratios vs energy, have the maximum (the absolute magnitude for $\pi$) at the same energy of 3.526 GeV, which possibly is the critical energy of phase transition from a liquid-like hadron state to a gas-like quark state in the collision system. With the increase of energy, all types of chemical potentials become small and tend to zero at very high energy, which indicates that with the increase of energy, the hadronic interactions gradually fade and the partonic interactions gradually become greater, and when the energy rises to a very high value, especially to the LHC, the collision system possibly changes completely from the liquid-like hadron-dominant state to the gas-like quark-dominant state, when the partonic interactions possibly play a dominant role.
\\

\\


\vskip1.0cm

{\section{Introduction}}

The successfully running of the Relativistic Heavy Ion Collider (RHIC) in 2000 and the Large Hadron Collider (LHC) in 2008, attracts great interest in studying the evolution process of interacting system in high energy collision. Many evidences confirm that such a high-energy collision system produces an extremely high  temperature and high density environment, which makes the collision system experience the phase transition process from the hadronic matter to quark-gluon plasma (QGP) or quark matter, and produce strong coupling quark-gluon plasma (sQGP). By studying the chemical freeze-out temperature ($T_{ch}$) of interacting system and the chemical potentials ($\mu_{B}$) of baryon in the phase diagram of quantum chromodynamics (QCD), one can obtain the information about the phase transition from hadronic matter to QGP or quark matter and the property of QGP, such as the critical end point (CEP) of phase transition. Thus, it's important to study baryon chemical potential in $\mu_{B}$-$T_{ch}$ plane. Meanwhile, the chemical potentials of other particles, like light hadrons and quarks, are also important and interesting in researching the evolution of collision system, the mechanism of particle production, and even the property of QGP.

The final-state particles produced in high energy collision are multifarious, and show many statistical behaviors which contain some information about collision process. It is interesting to find some useful information from these regular behaviors. After the kinetic freeze-out process of collision system, the transverse momentum of particle is no longer changing. So by analyzing transverse momentum distributions of final-state particles, one can obtain some information about the stage of kinetic freeze-out, even other stages of collision system. For example, one can extract the kinetic freeze-out temperature of interacting system , the flow velocity of particles  and so on directly from transverse momentum distributions. One can also extract the chemical potential of particles based on transverse momentum spectra and yield ratios of negative to positive particles at the stage of chemical freezing-out. Meanwhile, one can also  analyze the connections between these quantities and collision size, energy, centrality, particle mass, and so on, then further extract particle production mechanism and other information of other earlier stages.

Generally, one can use phenomenological model to describe the transverse momentum spectrum of final-state particles. These models can be divided into microcosmic kinetics model and thermal statistical model. Thermal statistical model focuses on studying the collective or global statistical behavior of final-state particles. There are many theoretical distribution models in the framework of thermal statistical model, such as Boltzmann distribution, blast-wave modle, power-law function, L$\acute{e}$vy distribution, Erlang distribution and so on. In the framework of multi-source thermal model, one can use multi-component distribution model to improve the fitting degree of single-component distribution in high transverse momentum region. Meanwhile, more information can be extracted. For example, by using multi-component Erlang distribution, one can extract not only the relative yield of particles, but also the weight of hard (soft) excitation degree.

By the yield ratios of negative to positive particles, one can obtain the chemical potentials of hadrons and quarks at the stage of chemical freeze-out according to references. While the yield ratios calculated from transverse momentum spectra of final-state particles, are actually at the stage of kinetic freeze-out, when the yield ratios are affected by the strong decay from high-mass resonance and the weak decay from heavy flavor hadrons. In order to obtain the yield ratios at the stage of chemical freeze-out, the contributions of strong decay and weak decay need to be removed from the yield ratios calculated from transverse momentum spectra. While according to the reference, the strong and weak decays are actually have less effect on the above particle yield ratios from normalization constants, although they have a big impact on particles yields. So, we can actually extract the chemical potentials of hadrons and quarks by using the yield ratios from normalization constants instead of the yield ratios modified by removing the contributions of strong and weak decays.

In the present work, we describe the transverse momentum ($p_{T}$) spectra of $\pi^\pm$, $K^\pm$, $p$, and $\bar{p}$ produced in different centralities gold-gold (Au-Au) collisions over a center-of-mass energy ($\sqrt{s_{NN}}$) range from 7.7 to 62.4 GeV by using a two-component Erlang distribution  in the framework of a multi-source thermal mode, and obtain the energy- and centrality-dependent yield ratios of negative to positive particles according to the extracted normalization constants. Meanwhile, the energy- and centrality-dependent chemical potentials of light hadrons ($\pi$, $K$, and $p$) and quarks ($u$, $d$, and $s$) are then extracted from the yield ratios.

{\section{The model and formulism}}

In the present work, we use a two-component Erlang distribution to describe the transverse momentum spectra of final-state light flavour particles to obtain the normalization constants, and to extract the yield ratios further. The two-component Erlang distribution is regarded as the contributions of the soft excitation process and the hard scattering process. The soft excitation process comes from the interactions among a few sea quarks and gluons and results in the low-$p_{T}$ region distribution, and the hard scattering process originates from a harder head-on scattering between two valent quarks and results in the high-$p_{T}$ region distribution. The two-component distribution is in the framework of a multi-source thermal model and the method is as follows.

The multi-source thermal model assumes that many emission sources are formed in high energy collisions. Due to the existent of different interacting mechanisms in the collisions and different event samples in experiment measurements, these emission sources are classified into $l$ groups. According to thermodynamic system, the transverse momentum of particles generated from one emission source obey to an exponential distribution,
\begin{equation}
f_{ij}(p_{tij})=\frac{1}{\langle{p_{tij}}\rangle}\exp{\bigg[-\frac{p_{tij}}{\langle{p_{tij}}\rangle}\bigg]},
\end{equation}
where $p_{tij}$ and $\langle{p_{tij}}\rangle$ are the transverse momentum of particles from the $i$-th source in the $j$-th group and the mean value of $p_{tij}$, respectively. Assume that the mean transverse momentum of particles from each source in the same group is the same. Then, all the sources in the $j$-th group meet the distribution of the folding result of exponential distribution
\begin{equation}
f_{j}(p_{T})=\frac{p_T^{m_{j}-1}}{(m_{j}-1)!\langle{p_{tij}}\rangle^{m_{j}}}\exp{\bigg[-\frac{p_{T}}{\langle{p_{tij}}\rangle}\bigg]},
\end{equation}
where $m_{j}$ is the source number in the $j$-th group and $p_{T}$ denotes the transverse momentum of particles from $m_{j}$ sources, i.e.
\begin{equation}
p_{T}=\sum_{i=1}^{m_{j}}p_{tij}.
\end{equation}
This is the normalized Erlang distribution, which can describe the $p_{T}$ distribution of the particles from the sources in the same group because they have the same excitation degree and stay at a common local equilibrium state. The contribution of all emission sources in all groups can be expressed as
\begin{equation}
f(p_{T})=\sum_{j=1}^{l}k_{j}f_{j}(p_{T}),
\end{equation}
where $k_{j}$ is the relative weight of the $j$-th group sources and meets the normalization $\sum_{j=1}^{l}k_{j}=1$. This is the multi-component Erlang distribution, which can describe the final-state $p_{T}$ distribution. Then, the two-component Erlang $p_{T}$ distribution can be written as
\begin{equation}
f(p_{T})=k_{1}f_{1}(p_{T})+(1-k_{1})f_{2}(p_{T}).
\end{equation}

According to the above model, we describe the $p_{T}$ spectra of $\pi^\pm$, $K^\pm$, $p$, and $\bar{p}$ produced in Au-Au collisions at different energies for differen centralities, and obtain the normalization constants corresponding to the above particles. The ratios of normalization constants of antiparticles, 
$\pi^{-}$, $K^{-}$, and $\overline{p}$, to particles, $\pi^{+}$, $K^{+}$, and $p$, are the yield ratios of negative to positive particles at the stage of kinetic freeze-out. Neglecting the effect of the strong and weak decays for the little contribution of the two decays to the yield ratios, the ratios of normalization constants are approximately equal to the yield ratios of particles at the stage of chemical freeze-out.

Based on the above yield ratios, we calculate the chemical potentials of some light hadrons ($\pi$, $K$, and $p$), and light quarks ($u$, $d$, and $s$). According to the statistical arguments based on the chemical and thermal equilibrium within the thermal and statistical model, the three types of yield ratios, $k_{\pi}$, $k_{K}$, and $k_{p}$, in terms of the light hadron chemical potentials, $\mu_{\pi}$, $\mu_{K}$, and $\mu_{p}$, of hadrons, $\pi$, $K$, and $p$, are to be 
\begin{gather}
k_{\pi}=\exp\bigg(-\frac{2\mu_{\pi}}{T_{ch}}\bigg),\notag\\
k_{K}=\exp\bigg(-\frac{2\mu_{K}}{T_{ch}}\bigg),\notag\\
k_{p}=\exp\bigg(-\frac{2\mu_{p}}{T_{ch}}\bigg),
\end{gather}
where $T_{ch}$ is the chemical freeze-out temperature of interacting system. Within the framework of a statistical thermal model of non-interacting gas particles with the assumption of standard Maxwell-Boltzmann statistics, $T_{ch}$ can be empirically obtained by the following formula
\begin{equation}
T_{ch}=T_{\lim}\frac{1}{1+\exp[2.60-\ln(\sqrt{s_{NN}})/0.45]},
\end{equation}
where $T_{\lim}$ is the `limiting' temperature and can be empirically taken the value of 0.164 GeV, and $\sqrt{s_{NN}}$ is in the unit of GeV.

Based on Equation (6) and references under the same value of chemical freeze-out temperature for $\pi$, $K$, and $p$, we can obtain the three types of yield ratios in terms of three types of quark chemical potentials ($\mu_{u}$, $\mu_{d}$, and $\mu_{s}$ for $u$, $d$, and $s$ quarks, respectively) to be

\begin{gather}
k_{\pi}=\exp\bigg[-\frac{(\mu_{u}-\mu_{d})}{T_{ch}}\bigg]\bigg/\exp\bigg[\frac{(\mu_{u}-\mu_{d})}{T_{ch}}\bigg]=\exp\bigg[-\frac{2(\mu_{u}-\mu_{d})}{T_{ch}}\bigg],\notag\\
k_{K}=\exp\bigg[-\frac{(\mu_{u}-\mu_{s})}{T_{ch}}\bigg]\bigg/\exp\bigg[\frac{(\mu_{u}-\mu_{s})}{T_{ch}}\bigg]=\exp\bigg[-\frac{2(\mu_{u}-\mu_{s})}{T_{ch}}\bigg],\notag\\
k_{p}=\exp\bigg[-\frac{(2\mu_{u}+\mu_{d})}{T_{ch}}\bigg]\bigg/\exp\bigg[\frac{(2\mu_{u}+\mu_{d})}{T_{ch}}\bigg]=\exp\bigg[-\frac{2(2\mu_{u}+\mu_{d})}{T_{ch}}\bigg].
\end{gather}

According to Equations (6) and (8), the chemical potentials of the above hadrons and quarks in terms of yield ratios can be respectively expressed as
\begin{gather}
\mu_{\pi}=-\frac{1}{2}T_{ch}\cdot\ln(k_{\pi}),\notag\\
\mu_{K}=-\frac{1}{2}T_{ch}\cdot\ln(k_{K}),\notag\\
\mu_{p}=-\frac{1}{2}T_{ch}\cdot\ln(k_{p}),
\end{gather}
and
\begin{gather}
\mu_{u}=-\frac{1}{6}T_{ch}\cdot\ln(k_{\pi}\cdot k_{p}),\notag\\
\mu_{d}=-\frac{1}{6}T_{ch}\cdot\ln(k_{\pi}^{-2}\cdot k_{p}),\notag\\
\mu_{s}=-\frac{1}{6}T_{ch}\cdot\ln(k_{\pi}\cdot k_{K}^{-3}\cdot
k_{p}).
\end{gather}

In the present work, we only calculate the chemical potentials of the light hadrons of $\pi$, $K$, and $p$, and light quarks of $u$, $d$, and $s$. For the hadrons containing $c$ or $b$ quark, considering that there is a lack of the experimental data of $p_{T}$ spectra which continuously vary with energy or centrality, we do not calculate the chemical potentials of $c$ and $b$ quarks, and the hadrons containing $c$ or $b$ quark. In addition, due to the lifetimes of the hadrons containing $t$ quark being too short to measure, we also can not obtain the chemical potentials of $t$ quark, and the hadrons containing $t$ quark.

{\section{Results and discussion}}

Figure 1 shows the transverse momentum distributions of (a)(d) $\pi^{\pm}$, (b)(e) $K^{\pm}$, (c) $p$, and (f) $\bar{p}$ produced in Au-Au collisions at $\sqrt{s_{NN}}=$ 7.7 GeV in different centrality classes of 0--5\%, 5--10\%, 10--20\%, 20--30\%, 30--40\%, 40--50\%, 50--60\%, 60--70\%, and 70--80\%. The transverse momentum spectra for the energies of 11.5, 19.6, 27, 39, and 62.4 GeV are presented in Figures 2--6, respectively. $dN/dy$ on axis denotes the rapidity density. The symbols represent the experimental data recorded by the STAR Collaboration in the mid-rapidity range of $|y|<0.1$. The spectra for all centralities are scaled by suitable factors for clarity. The uncertainties are statistical and systematic added in quadrature. The curves are our results calculated by using the two-component Erlang distribution. The values of free parameters ($m_{1}$, $p_{ti1}$, $k_{1}$, $m_{2}$, and $p_{ti2}$), normalization constant ($N_0$), and $\chi^2$ per degree of freedom ($\chi^2$/dof) corresponding to the two-component Erlang distribution for different energies are respectively listed in Tables 1--6, where the normalization constant is for comparison between curve and data. One can see that the two-component Erlang distribution can well describe the experimental data of the considered particles in Au-Au collisions at all energies for all centrality classes. The tables show that the values of $m_{1}$ corresponding to low-$p_{T}$ region for all particles at all energies in all centrality classes are 2, 3, or 4, and all $m_{2}$ corresponding to high-$p_{T}$ region are 2, which shows that the soft process origins from the interaction among 2, 3, or 4 sea quarks and gluons, and the hard process origins from a hard head-on scattering between two valent quarks. The values of relative weight factor $k_1$ of soft excitation process are more than 60\%, which reflects that the soft excitation is the main excitation process. In addition, the normalization constant $N_0$ increases with increase of energy and centrality, and decreases with increase of particle mass.

\begin{figure}[H]
\hskip-0.0cm {\centering
\includegraphics[width=16.0cm]{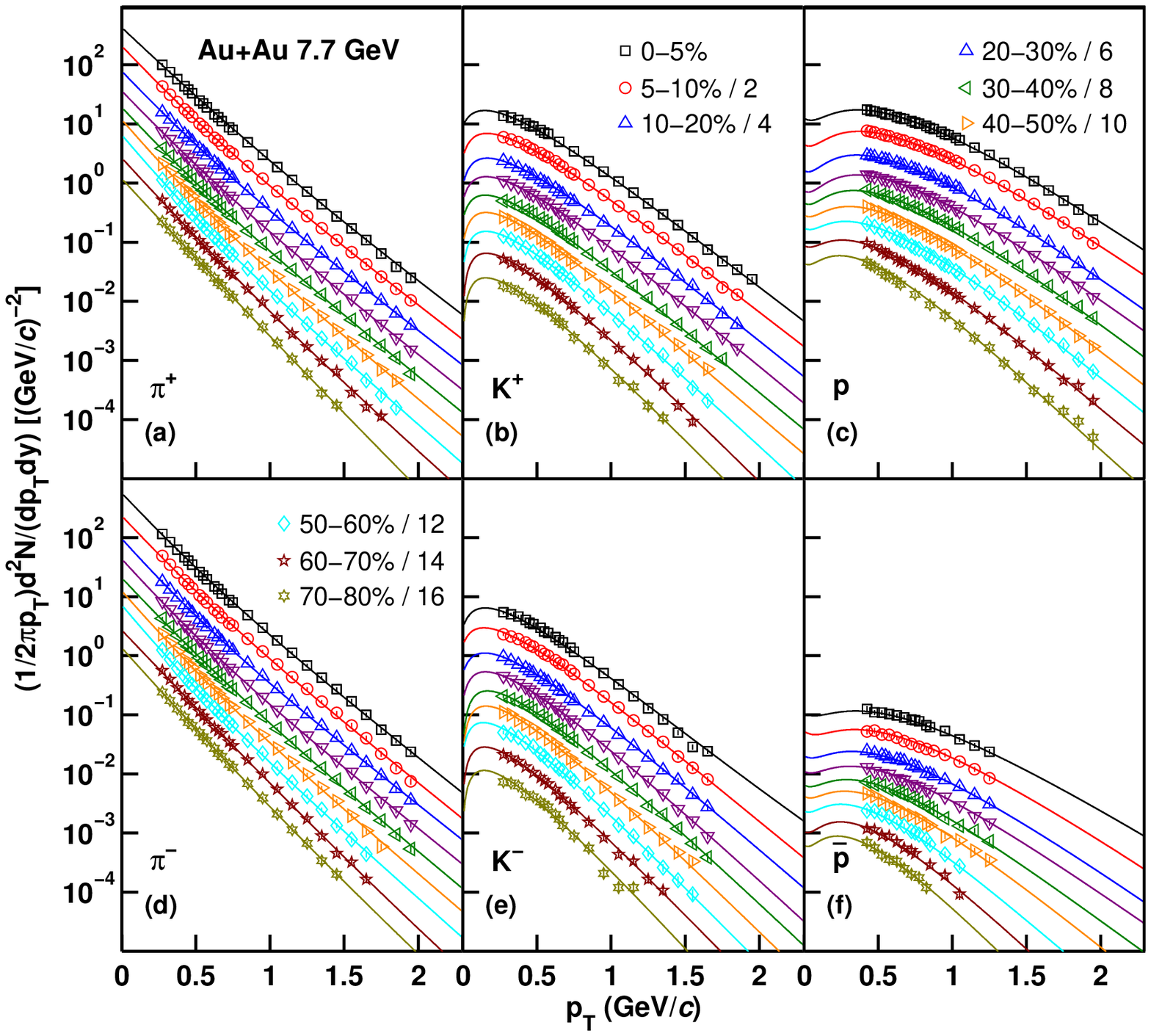}}
\vskip-0.18cm Figure 1. Transverse momentum spectra for (a)(b)(c) positive ($\pi^+$, $K^+$, $p$) and (d)(e)(f) negative ($\pi^-$, $K^-$, $\bar{p}$) particles produced in Au-Au collisions with $|y|<0.1$ at $\sqrt{s_{NN}}=$ 7.7 GeV for different centralities (0--5\%, 5--10\%, 10--20\%, 20--30\%, 30--40\%, 40--50\%, 50--60\%, 60--70\%, and 70--80\%). The experimental data represented by the symbols are measured by the STAR Collaboration. The spectra for different centralities are scaled by suitable factors for clarity. The plotted errors bars include both statistical and systematic uncertainties, and the curves are the two-component Erlang distribution fits to the spectra.
\end{figure}

\begin{figure}[H]
\hskip-0.0cm {\centering
\includegraphics[width=16.0cm]{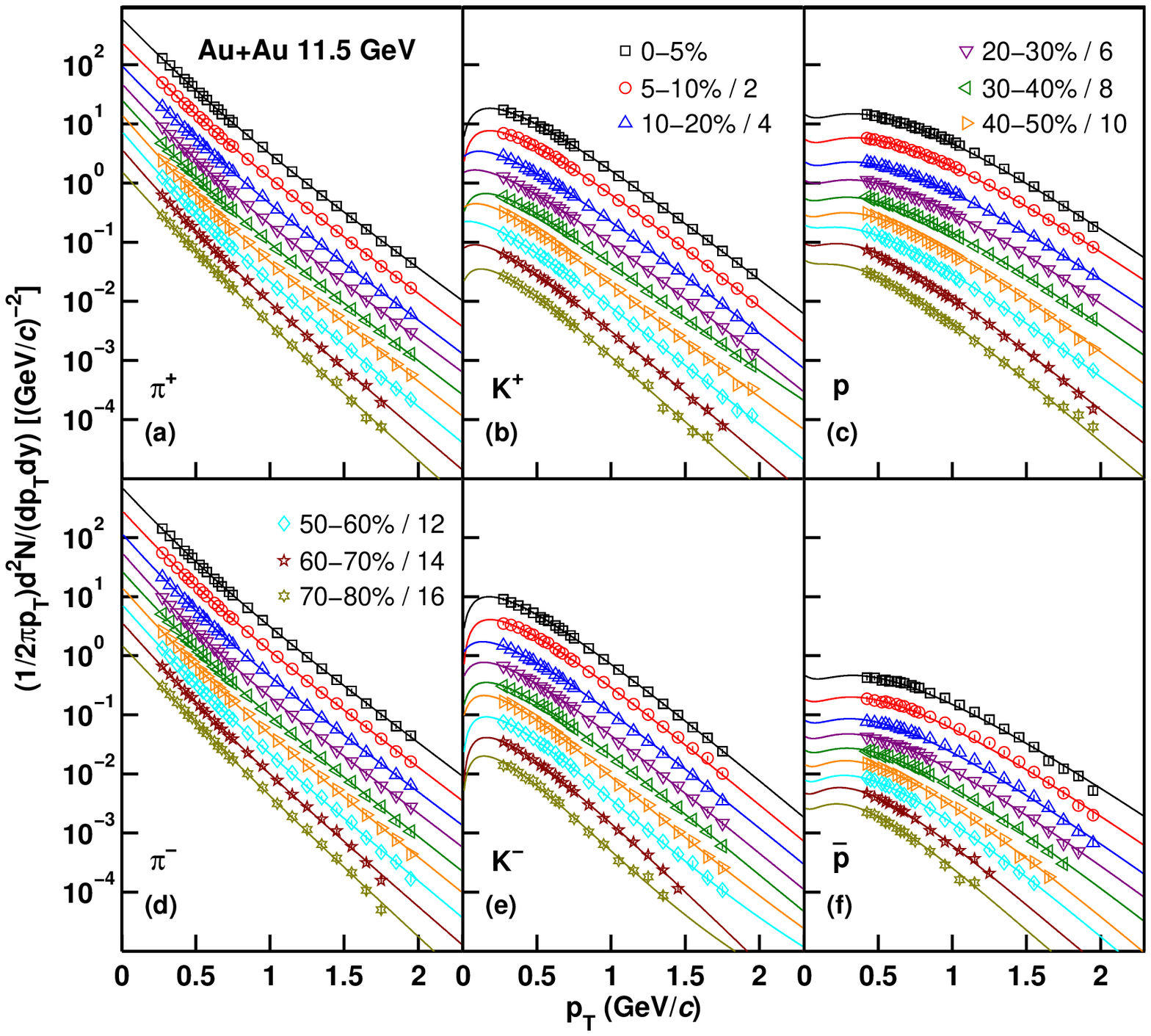}}
\vskip-0.18cm  Figure 2. Same as Figure 1 but for Au-Au collisions at $\sqrt{s_{NN}}=$ 11.5 GeV.
\end{figure}

\begin{figure}[H]
\hskip-0.0cm {\centering
\includegraphics[width=16.0cm]{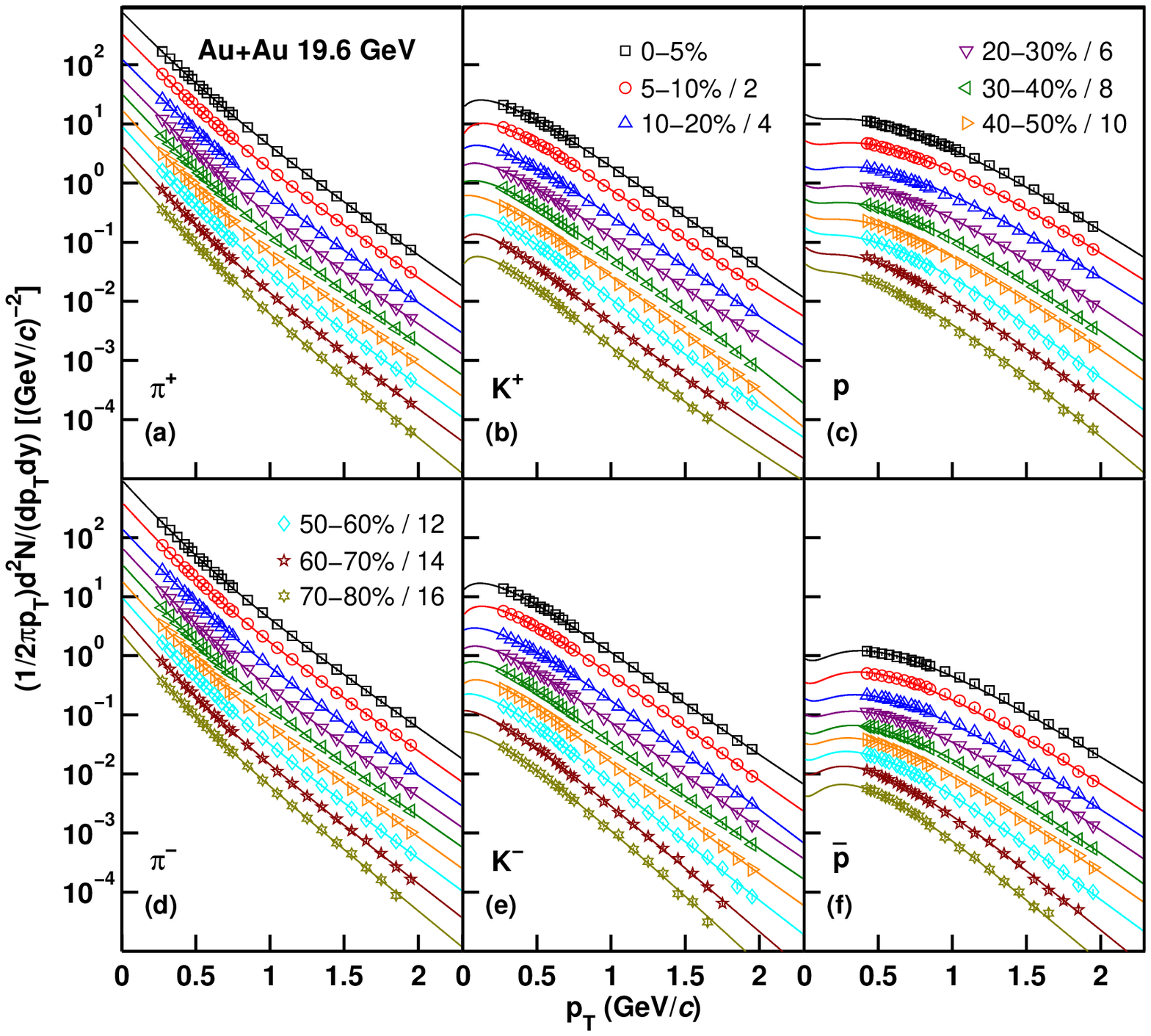}}
\vskip-0.18cm  Figure 3. Same as Figure 1 but for Au-Au collisions at $\sqrt{s_{NN}}=$ 19.6 GeV.
\end{figure}

\begin{figure}[H]
\hskip-0.0cm {\centering
\includegraphics[width=16.0cm]{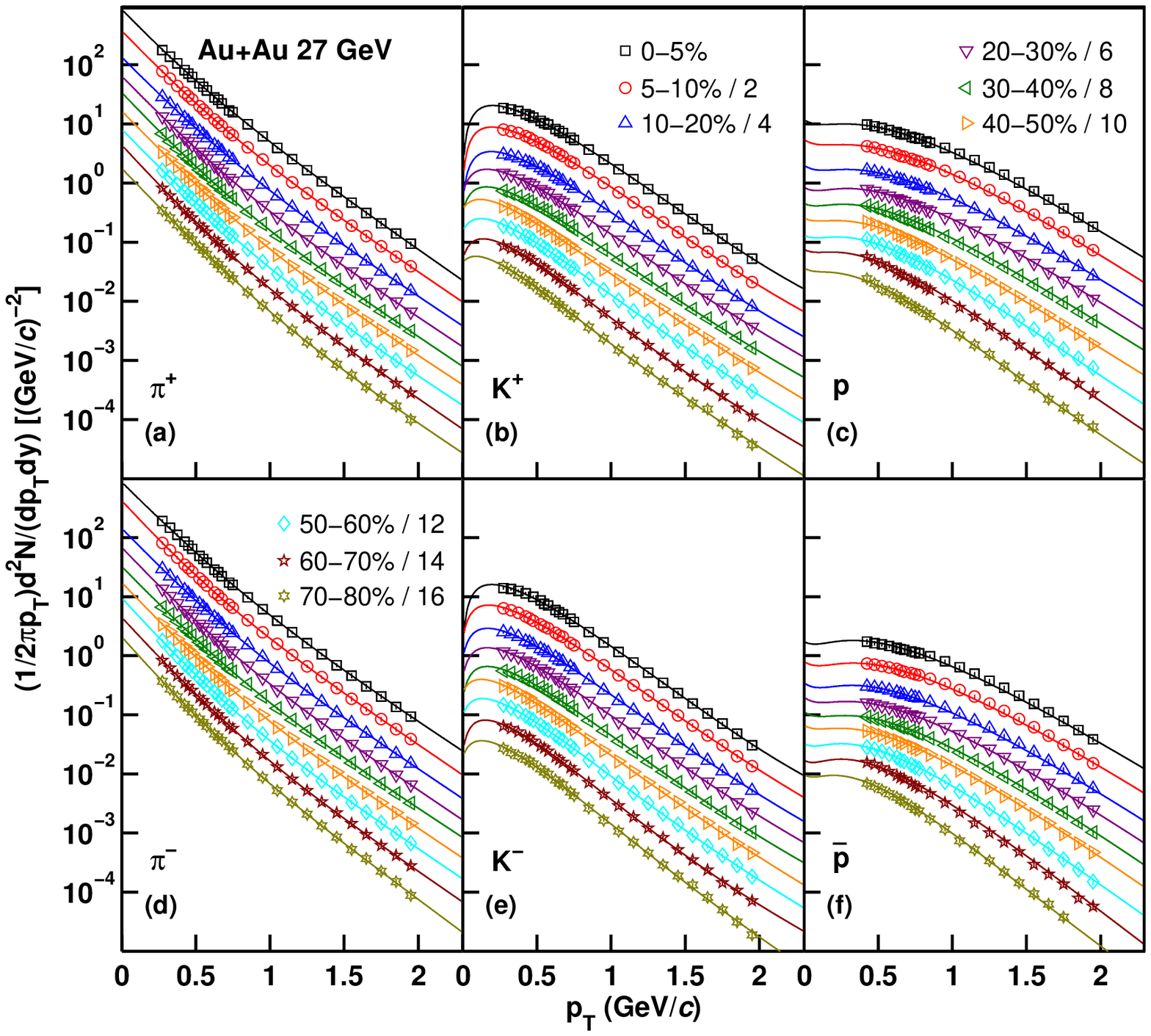}}
\vskip-0.18cm  Figure 4. Same as Figure 1 but for Au-Au collisions at $\sqrt{s_{NN}}=$ 27 GeV.
\end{figure}

\begin{figure}[H]
\hskip-0.0cm {\centering
\includegraphics[width=16.0cm]{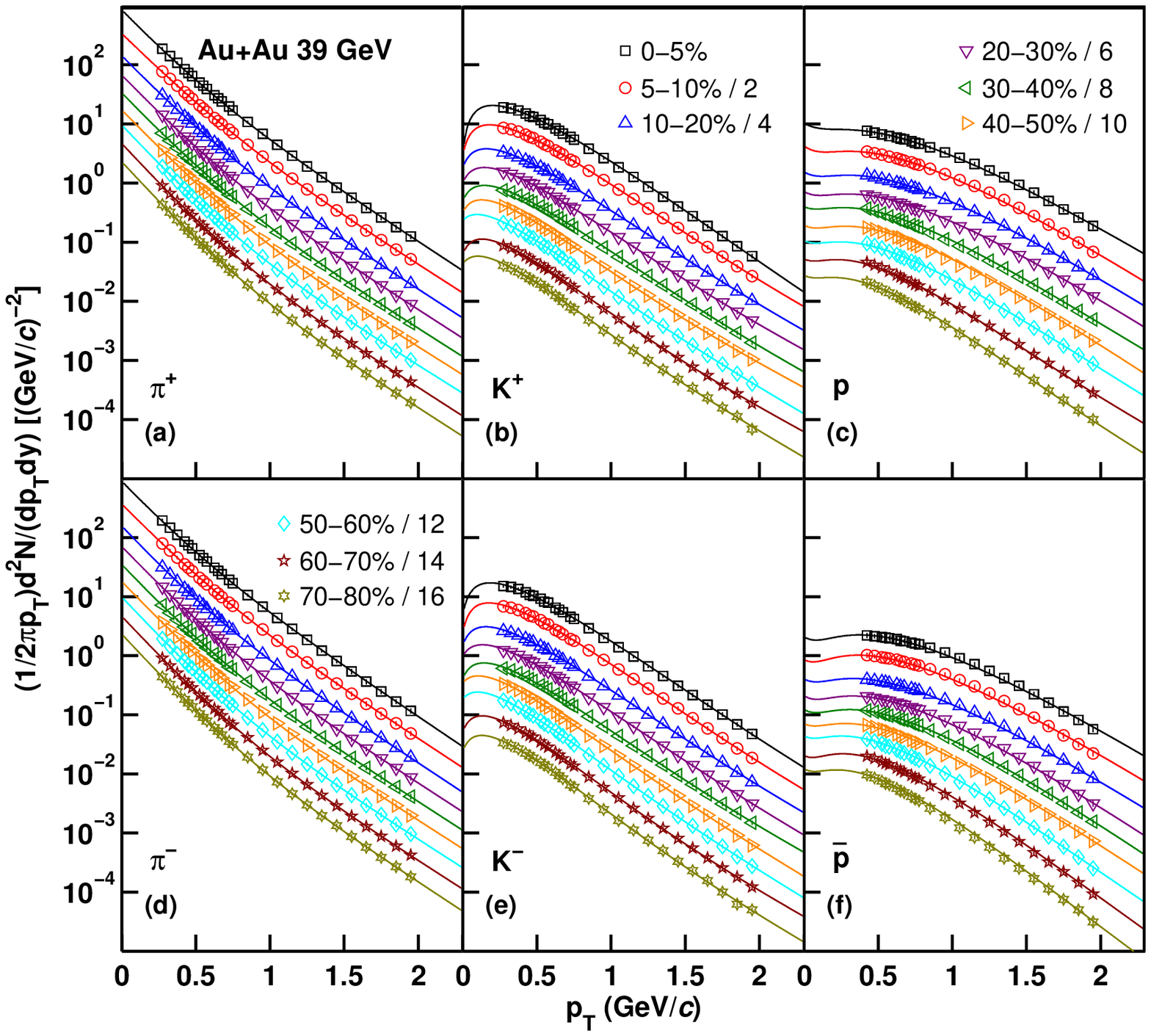}}
\vskip-0.18cm  Figure 5. Same as Figure 1 but for Au-Au collisions at $\sqrt{s_{NN}}=$ 39 GeV.
\end{figure}

\begin{figure}[H]
\hskip-0.0cm {\centering
\includegraphics[width=16.0cm]{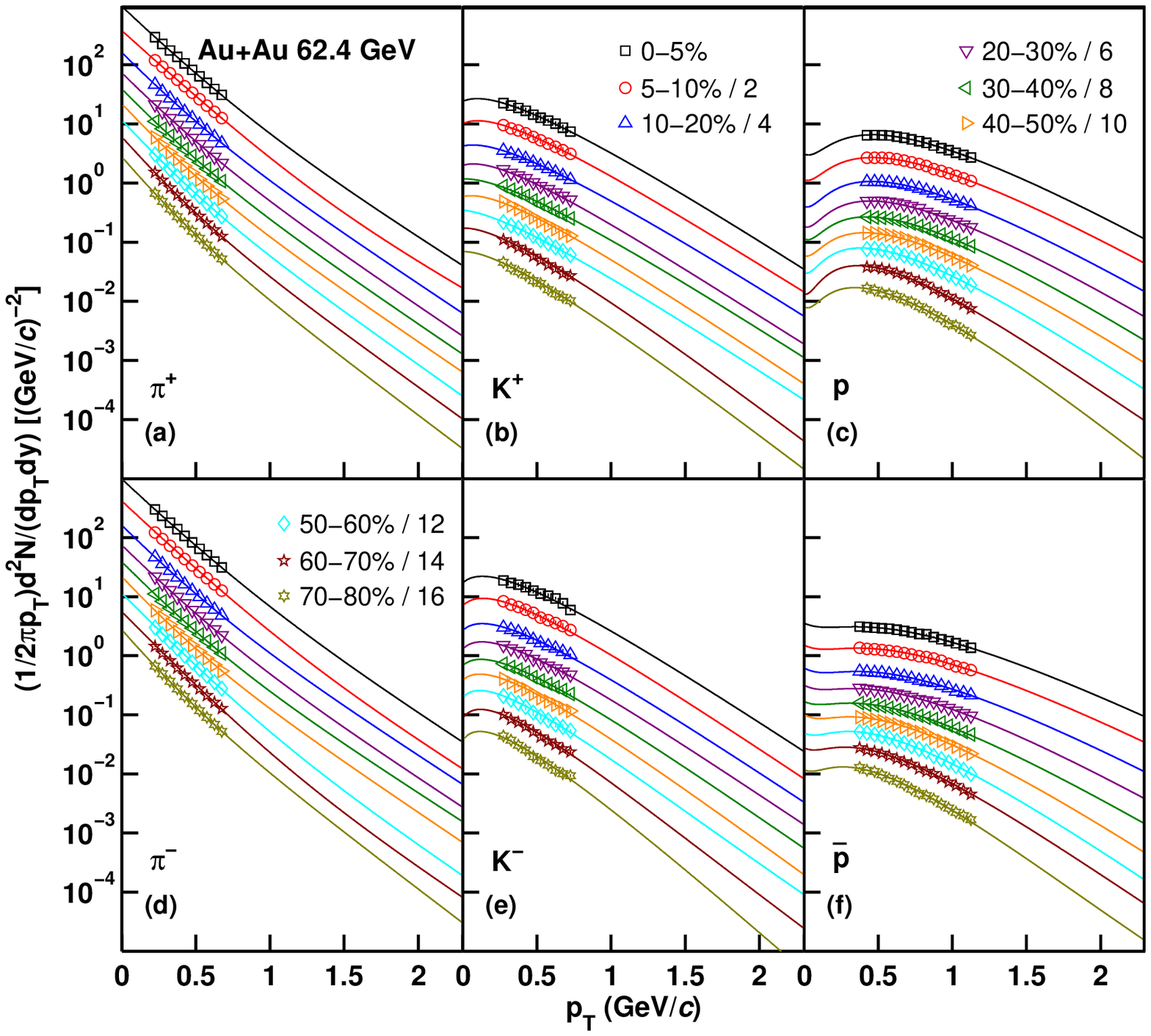}}
\vskip-0.18cm  Figure 6. Same as Figure 1 but for Au-Au collisions at $\sqrt{s_{NN}}=$ 62.4 GeV. The experimental data were recorded by the STAR Collaboration.
\end{figure}

\newpage

{\scriptsize {Table 1. Values of free parameters, normalization constant, and $\chi^2$/dof  corresponding to the two-component Erlang $p_T$ distribution for productions in Au+Au collisions at $\sqrt{s_{NN}}=\rm{7.7\, GeV}$ for different centralities in Figure 1.
{%
\begin{center}
\begin{tabular}{ccccccccccc}
\hline
\hline
Figure& Particle& Centrality & $m_1$ &$<p_{ti1}>$ & $k_1$ & $m_2$ &$<p_{ti2}>$ & $N_0$& $\chi^2$/dof \\
 &  (GeV) &  &   & (GeV/c) &  & & (GeV/c)& & \\
\hline
&  & 0-5\% & 2 &    0.172 $\pm$    0.004 &     0.63 $\pm$     0.06 & 2 &    0.233 $\pm$    0.004 &   96.122 $\pm$    3.364 &   13.970/20\\
&  & 5-10\% & 2 &    0.148 $\pm$    0.006 &     0.51 $\pm$     0.03 & 2 &    0.225 $\pm$    0.002 &   80.050 $\pm$    2.642 &    3.319/20\\
&  & 10-20\% & 2 &    0.150 $\pm$    0.006 &     0.51 $\pm$     0.04 & 2 &    0.224 $\pm$    0.003 &   60.889 $\pm$    2.192 &    1.896/20\\
&  & 20-30\% & 2 &    0.150 $\pm$    0.006 &     0.51 $\pm$     0.03 & 2 &    0.219 $\pm$    0.002 &   41.772 $\pm$    1.378 &    3.579/20\\
Figure 1(a)&  $\pi^{+}$ & 30-40\% & 2 &    0.147 $\pm$    0.006 &     0.51 $\pm$     0.03 & 2 &    0.214 $\pm$    0.003 &   28.090 $\pm$    0.955 &    5.519/20\\
&  & 40-50\% & 2 &    0.135 $\pm$    0.005 &     0.51 $\pm$     0.03 & 2 &    0.207 $\pm$    0.003 &   18.994 $\pm$    0.646 &    3.120/19\\
&  & 50-60\% & 2 &    0.133 $\pm$    0.006 &     0.51 $\pm$     0.03 & 2 &    0.198 $\pm$    0.003 &   11.956 $\pm$    0.383 &    8.153/19\\
&  & 60-70\% & 2 &    0.148 $\pm$    0.004 &     0.59 $\pm$     0.05 & 2 &    0.195 $\pm$    0.003 &    6.225 $\pm$    0.212 &    9.046/18\\
&  & 70-80\% & 2 &    0.151 $\pm$    0.004 &     0.53 $\pm$     0.07 & 2 &    0.176 $\pm$    0.004 &    3.125 $\pm$    0.103 &    3.702/15\\
\hline
&  & 0-5\% & 2 &    0.149 $\pm$    0.006 &     0.52 $\pm$     0.03 & 2 &    0.219 $\pm$    0.003 &  107.122 $\pm$    3.642 &    9.297/20\\
&  & 5-10\% & 2 &    0.147 $\pm$    0.006 &     0.51 $\pm$     0.03 & 2 &    0.216 $\pm$    0.002 &   86.526 $\pm$    3.028 &   16.388/20\\
&  & 10-20\% & 2 &    0.141 $\pm$    0.007 &     0.51 $\pm$     0.03 & 2 &    0.219 $\pm$    0.002 &   67.409 $\pm$    2.494 &    3.500/20\\
&  & 20-30\% & 2 &    0.142 $\pm$    0.006 &     0.51 $\pm$     0.03 & 2 &    0.215 $\pm$    0.003 &   45.954 $\pm$    1.608 &    5.489/20\\
Figure 1(d)&  $\pi^{-}$ & 30-40\% & 2 &    0.147 $\pm$    0.007 &     0.51 $\pm$     0.03 & 2 &    0.209 $\pm$    0.003 &   30.232 $\pm$    1.028 &    7.770/20\\
&  & 40-50\% & 2 &    0.137 $\pm$    0.006 &     0.51 $\pm$     0.03 & 2 &    0.203 $\pm$    0.003 &   20.251 $\pm$    0.689 &    9.656/18\\
&  & 50-60\% & 2 &    0.132 $\pm$    0.006 &     0.51 $\pm$     0.03 & 2 &    0.195 $\pm$    0.002 &   12.950 $\pm$    0.414 &    7.004/17\\
&  & 60-70\% & 2 &    0.161 $\pm$    0.003 &     0.83 $\pm$     0.05 & 2 &    0.200 $\pm$    0.006 &    6.606 $\pm$    0.231 &    9.005/17\\
&  & 70-80\% & 2 &    0.146 $\pm$    0.004 &     0.61 $\pm$     0.05 & 2 &    0.185 $\pm$    0.004 &    3.438 $\pm$    0.131 &    2.791/15\\
\hline
&  & 0-5\% & 3 &    0.197 $\pm$    0.003 &     0.81 $\pm$     0.16 & 2 &    0.260 $\pm$    0.018 &   20.070 $\pm$    0.682 &    9.050/17\\
&  & 5-10\% & 3 &    0.193 $\pm$    0.003 &     0.86 $\pm$     0.15 & 2 &    0.270 $\pm$    0.024 &   16.565 $\pm$    0.514 &    6.988/19\\
&  & 10-20\% & 3 &    0.191 $\pm$    0.002 &     0.89 $\pm$     0.17 & 2 &    0.255 $\pm$    0.017 &   12.444 $\pm$    0.373 &   11.722/19\\
&  & 20-30\% & 3 &    0.186 $\pm$    0.002 &     0.88 $\pm$     0.15 & 2 &    0.244 $\pm$    0.016 &    8.508 $\pm$    0.264 &    6.350/19\\
Figure 1(b)& $K^{+}$ & 30-40\% & 3 &    0.178 $\pm$    0.003 &     0.86 $\pm$     0.16 & 2 &    0.252 $\pm$    0.016 &    5.111 $\pm$    0.164 &   13.359/18\\
&  & 40-50\% & 3 &    0.173 $\pm$    0.003 &     0.90 $\pm$     0.18 & 2 &    0.255 $\pm$    0.043 &    3.150 $\pm$    0.101 &    7.742/17\\
&  & 50-60\% & 3 &    0.168 $\pm$    0.002 &     0.94 $\pm$     0.14 & 2 &    0.231 $\pm$    0.032 &    1.721 $\pm$    0.053 &    5.944/16\\
&  & 60-70\% & 3 &    0.161 $\pm$    0.002 &     0.99 $\pm$     0.15 & 2 &    0.202 $\pm$    0.040 &    0.800 $\pm$    0.034 &    8.731/15\\
&  & 70-80\% & 3 &    0.156 $\pm$    0.003 &     0.99 $\pm$     0.18 & 2 &    0.230 $\pm$    0.046 &    0.330 $\pm$    0.013 &   13.035/12\\
\hline
&  & 0-5\% & 3 &    0.186 $\pm$    0.004 &     0.80 $\pm$     0.16 & 2 &    0.285 $\pm$    0.015 &    7.208 $\pm$    0.252 &   40.240/17\\
&  & 5-10\% & 3 &    0.182 $\pm$    0.003 &     0.82 $\pm$     0.16 & 2 &    0.247 $\pm$    0.018 &    6.131 $\pm$    0.196 &   10.748/17\\
&  & 10-20\% & 3 &    0.182 $\pm$    0.002 &     0.87 $\pm$     0.16 & 2 &    0.226 $\pm$    0.024 &    4.620 $\pm$    0.143 &    4.469/17\\
&  & 20-30\% & 3 &    0.174 $\pm$    0.002 &     0.93 $\pm$     0.10 & 2 &    0.195 $\pm$    0.039 &    3.129 $\pm$    0.113 &    4.821/17\\
Figure 1(e)& $K^{-}$ & 30-40\% & 3 &    0.170 $\pm$    0.002 &     0.99 $\pm$     0.10 & 2 &    0.204 $\pm$    0.040 &    1.990 $\pm$    0.070 &    6.867/17\\
&  & 40-50\% & 3 &    0.162 $\pm$    0.002 &     0.97 $\pm$     0.19 & 2 &    0.220 $\pm$    0.044 &    1.243 $\pm$    0.044 &    8.169/14\\
&  & 50-60\% & 3 &    0.155 $\pm$    0.002 &     0.92 $\pm$     0.14 & 2 &    0.195 $\pm$    0.034 &    0.692 $\pm$    0.027 &   12.526/15\\
&  & 60-70\% & 3 &    0.152 $\pm$    0.003 &     0.97 $\pm$     0.19 & 2 &    0.218 $\pm$    0.043 &    0.312 $\pm$    0.011 &   12.456/13\\
&  & 70-80\% & 3 &    0.146 $\pm$    0.004 &     0.99 $\pm$     0.19 & 2 &    0.180 $\pm$    0.036 &    0.133 $\pm$    0.006 &   27.764/10\\
\hline
&  & 0-5\% & 4 &    0.215 $\pm$    0.003 &     0.89 $\pm$     0.08 & 2 &    0.270 $\pm$    0.054 &   52.211 $\pm$    2.193 &    4.927/23\\
&  & 5-10\% & 4 &    0.211 $\pm$    0.003 &     0.91 $\pm$     0.08 & 2 &    0.265 $\pm$    0.053 &   44.223 $\pm$    1.946 &    3.353/23\\
&  & 10-20\% & 4 &    0.201 $\pm$    0.003 &     0.91 $\pm$     0.09 & 2 &    0.265 $\pm$    0.053 &   32.020 $\pm$    1.473 &    9.462/23\\
&  & 20-30\% & 4 &    0.200 $\pm$    0.003 &     0.92 $\pm$     0.10 & 2 &    0.250 $\pm$    0.050 &   21.932 $\pm$    0.855 &    2.729/23\\
Figure 1(c)&  $p$ & 30-40\% & 4 &    0.192 $\pm$    0.003 &     0.90 $\pm$     0.09 & 2 &    0.250 $\pm$    0.050 &   14.565 $\pm$    0.612 &    4.083/22\\
&  & 40-50\% & 4 &    0.183 $\pm$    0.003 &     0.86 $\pm$     0.09 & 2 &    0.260 $\pm$    0.035 &    8.745 $\pm$    0.350 &    4.582/22\\
&  & 50-60\% & 4 &    0.174 $\pm$    0.003 &     0.83 $\pm$     0.14 & 2 &    0.260 $\pm$    0.035 &    5.248 $\pm$    0.210 &   18.492/21\\
&  & 60-70\% & 4 &    0.164 $\pm$    0.003 &     0.80 $\pm$     0.16 & 2 &    0.260 $\pm$    0.019 &    2.622 $\pm$    0.105 &   13.162/22\\
&  & 70-80\% & 4 &    0.148 $\pm$    0.002 &     0.81 $\pm$     0.16 & 2 &    0.240 $\pm$    0.016 &    1.345 $\pm$    0.059 &   29.998/15\\
\hline
&  & 0-5\% & 4 &    0.232 $\pm$    0.011 &     0.81 $\pm$     0.16 & 2 &    0.334 $\pm$    0.066 &    0.392 $\pm$    0.019 &    3.384/9\\
&  & 5-10\% & 4 &    0.217 $\pm$    0.009 &     0.88 $\pm$     0.14 & 2 &    0.247 $\pm$    0.049 &    0.338 $\pm$    0.016 &    7.632/8\\
&  & 10-20\% & 4 &    0.205 $\pm$    0.009 &     0.82 $\pm$     0.11 & 2 &    0.302 $\pm$    0.060 &    0.257 $\pm$    0.012 &    3.345/12\\
&  & 20-30\% & 4 &    0.198 $\pm$    0.009 &     0.80 $\pm$     0.16 & 2 &    0.310 $\pm$    0.062 &    0.202 $\pm$    0.009 &    9.058/10\\
Figure 1(f)&  $\overline{p}$ & 30-40\% & 4 &    0.184 $\pm$    0.008 &     0.79 $\pm$     0.15 & 2 &    0.300 $\pm$    0.060 &    0.140 $\pm$    0.007 &    3.650/11\\
&  & 40-50\% & 4 &    0.174 $\pm$    0.006 &     0.81 $\pm$     0.16 & 2 &    0.266 $\pm$    0.053 &    0.099 $\pm$    0.006 &    8.561/8\\
&  & 50-60\% & 4 &    0.156 $\pm$    0.007 &     0.79 $\pm$     0.15 & 2 &    0.260 $\pm$    0.052 &    0.058 $\pm$    0.003 &    5.627/7\\
&  & 60-70\% & 4 &    0.148 $\pm$    0.007 &     0.83 $\pm$     0.16 & 2 &    0.236 $\pm$    0.047 &    0.031 $\pm$    0.002 &    4.186/5\\
&  & 70-80\% & 4 &    0.137 $\pm$    0.011 &     0.82 $\pm$     0.16 & 2 &    0.224 $\pm$    0.044 &    0.017 $\pm$    0.002 &    1.189/3\\
\hline
\hline
\end{tabular}
\end{center}
}} }

\newpage

{\scriptsize {Table 2. Values of free parameters, normalization constant, and $\chi^2$/dof  corresponding to the two-component Erlang $p_T$ distribution for productions in Au+Au collisions at $\sqrt{s_{NN}}=\rm{11.5\, GeV}$ for different centralities in Figure 2.
{%
\begin{center}
\begin{tabular}{ccccccccccc}
\hline
\hline
Figure& Particle& Centrality & $m_1$ &$<p_{ti1}>$ & $k_1$ & $m_2$ &$<p_{ti2}>$ & $N_0$& $\chi^2$/dof \\
 &  (GeV) &  &   & (GeV/c) &  & & (GeV/c)& & \\
\hline
&  & 0-5\% & 2 &    0.153 $\pm$    0.007 &     0.52 $\pm$     0.04 & 2 &    0.236 $\pm$    0.003 &  125.208 $\pm$    4.633 &    2.186/20\\
&  & 5-10\% & 2 &    0.153 $\pm$    0.006 &     0.51 $\pm$     0.04 & 2 &    0.233 $\pm$    0.003 &   98.692 $\pm$    3.257 &    1.584/20\\
&  & 10-20\% & 2 &    0.149 $\pm$    0.007 &     0.51 $\pm$     0.03 & 2 &    0.230 $\pm$    0.003 &   76.333 $\pm$    3.053 &    1.429/20\\
&  & 20-30\% & 2 &    0.143 $\pm$    0.006 &     0.53 $\pm$     0.04 & 2 &    0.233 $\pm$    0.003 &   52.743 $\pm$    1.846 &    1.146/20\\
Figure 2(a)&  $\pi^{+}$ & 30-40\% & 2 &    0.141 $\pm$    0.006 &     0.53 $\pm$     0.04 & 2 &    0.227 $\pm$    0.003 &   36.192 $\pm$    1.231 &    0.998/20\\
&  & 40-50\% & 2 &    0.133 $\pm$    0.006 &     0.51 $\pm$     0.03 & 2 &    0.221 $\pm$    0.003 &   23.473 $\pm$    0.845 &    1.222/20\\
&  & 50-60\% & 2 &    0.132 $\pm$    0.006 &     0.51 $\pm$     0.03 & 2 &    0.212 $\pm$    0.002 &   14.263 $\pm$    0.571 &    2.065/20\\
&  & 60-70\% & 2 &    0.133 $\pm$    0.006 &     0.51 $\pm$     0.03 & 2 &    0.204 $\pm$    0.003 &    8.159 $\pm$    0.310 &    2.188/18\\
&  & 70-80\% & 2 &    0.141 $\pm$    0.006 &     0.52 $\pm$     0.04 & 2 &    0.197 $\pm$    0.003 &    4.135 $\pm$    0.141 &    5.479/18\\
\hline
&  & 0-5\% & 2 &    0.146 $\pm$    0.008 &     0.51 $\pm$     0.04 & 2 &    0.230 $\pm$    0.003 &  135.170 $\pm$    5.812 &    1.593/20\\
&  & 5-10\% & 2 &    0.143 $\pm$    0.007 &     0.51 $\pm$     0.04 & 2 &    0.229 $\pm$    0.002 &  107.425 $\pm$    4.082 &    1.383/20\\
&  & 10-20\% & 2 &    0.140 $\pm$    0.005 &     0.53 $\pm$     0.04 & 2 &    0.230 $\pm$    0.003 &   83.065 $\pm$    2.658 &    0.781/20\\
&  & 20-30\% & 2 &    0.135 $\pm$    0.006 &     0.51 $\pm$     0.04 & 2 &    0.227 $\pm$    0.003 &   56.923 $\pm$    1.992 &    0.895/20\\
Figure 2(d)&  $\pi^{-}$ & 30-40\% & 2 &    0.140 $\pm$    0.006 &     0.51 $\pm$     0.03 & 2 &    0.220 $\pm$    0.003 &   38.112 $\pm$    1.334 &    2.378/20\\
&  & 40-50\% & 2 &    0.137 $\pm$    0.007 &     0.51 $\pm$     0.03 & 2 &    0.216 $\pm$    0.003 &   24.354 $\pm$    0.950 &    4.048/20\\
&  & 50-60\% & 2 &    0.137 $\pm$    0.006 &     0.51 $\pm$     0.03 & 2 &    0.209 $\pm$    0.003 &   14.725 $\pm$    0.560 &    7.309/20\\
&  & 60-70\% & 2 &    0.139 $\pm$    0.005 &     0.51 $\pm$     0.03 & 2 &    0.201 $\pm$    0.003 &    8.447 $\pm$    0.287 &   11.989/18\\
&  & 70-80\% & 2 &    0.150 $\pm$    0.005 &     0.51 $\pm$     0.04 & 2 &    0.191 $\pm$    0.003 &    4.250 $\pm$    0.153 &   10.734/18\\
\hline
&  & 0-5\% & 3 &    0.201 $\pm$    0.003 &     0.93 $\pm$     0.18 & 2 &    0.262 $\pm$    0.048 &   24.436 $\pm$    0.733 &    1.486/19\\
&  & 5-10\% & 3 &    0.198 $\pm$    0.002 &     0.96 $\pm$     0.08 & 2 &    0.207 $\pm$    0.041 &   19.832 $\pm$    0.714 &    2.160/20\\
&  & 10-20\% & 3 &    0.200 $\pm$    0.002 &     0.82 $\pm$     0.06 & 2 &    0.199 $\pm$    0.026 &   14.781 $\pm$    0.473 &    2.322/20\\
&  & 20-30\% & 3 &    0.199 $\pm$    0.003 &     0.76 $\pm$     0.06 & 2 &    0.205 $\pm$    0.020 &    9.726 $\pm$    0.350 &    3.432/20\\
Figure 2(b)& $K^{+}$ & 30-40\% & 3 &    0.184 $\pm$    0.003 &     0.79 $\pm$     0.05 & 2 &    0.306 $\pm$    0.010 &    5.985 $\pm$    0.180 &    1.553/20\\
&  & 40-50\% & 3 &    0.174 $\pm$    0.003 &     0.59 $\pm$     0.09 & 2 &    0.260 $\pm$    0.007 &    3.861 $\pm$    0.131 &    5.020/20\\
&  & 50-60\% & 3 &    0.180 $\pm$    0.003 &     0.54 $\pm$     0.10 & 2 &    0.236 $\pm$    0.006 &    2.000 $\pm$    0.060 &    5.888/19\\
&  & 60-70\% & 3 &    0.172 $\pm$    0.003 &     0.65 $\pm$     0.13 & 2 &    0.228 $\pm$    0.009 &    0.997 $\pm$    0.033 &    2.330/17\\
&  & 70-80\% & 3 &    0.163 $\pm$    0.003 &     0.78 $\pm$     0.15 & 2 &    0.231 $\pm$    0.016 &    0.464 $\pm$    0.015 &   11.598/16\\
\hline
&  & 0-5\% & 3 &    0.191 $\pm$    0.003 &     0.96 $\pm$     0.11 & 2 &    0.221 $\pm$    0.044 &   12.017 $\pm$    0.385 &    1.338/17\\
&  & 5-10\% & 3 &    0.188 $\pm$    0.002 &     0.98 $\pm$     0.11 & 2 &    0.295 $\pm$    0.059 &    9.851 $\pm$    0.325 &    7.987/18\\
&  & 10-20\% & 3 &    0.193 $\pm$    0.003 &     0.83 $\pm$     0.16 & 2 &    0.219 $\pm$    0.022 &    7.509 $\pm$    0.225 &   10.112/18\\
&  & 20-30\% & 3 &    0.185 $\pm$    0.003 &     0.82 $\pm$     0.16 & 2 &    0.245 $\pm$    0.018 &    4.915 $\pm$    0.152 &    3.830/18\\
Figure 2(e)& $K^{-}$ & 30-40\% & 3 &    0.182 $\pm$    0.002 &     0.88 $\pm$     0.11 & 2 &    0.256 $\pm$    0.015 &    3.046 $\pm$    0.091 &    3.806/17\\
&  & 40-50\% & 3 &    0.167 $\pm$    0.003 &     0.82 $\pm$     0.06 & 2 &    0.275 $\pm$    0.010 &    1.974 $\pm$    0.069 &    4.685/17\\
&  & 50-60\% & 3 &    0.164 $\pm$    0.003 &     0.92 $\pm$     0.04 & 2 &    0.327 $\pm$    0.027 &    1.031 $\pm$    0.032 &    4.597/17\\
&  & 60-70\% & 3 &    0.162 $\pm$    0.003 &     0.99 $\pm$     0.16 & 2 &    0.254 $\pm$    0.050 &    0.517 $\pm$    0.019 &    5.856/14\\
&  & 70-80\% & 3 &    0.148 $\pm$    0.004 &     0.91 $\pm$     0.06 & 2 &    0.325 $\pm$    0.046 &    0.246 $\pm$    0.010 &   14.250/10\\
\hline
&  & 0-5\% & 4 &    0.213 $\pm$    0.004 &     0.88 $\pm$     0.08 & 2 &    0.234 $\pm$    0.046 &   42.924 $\pm$    1.803 &    8.894/22\\
&  & 5-10\% & 4 &    0.214 $\pm$    0.005 &     0.89 $\pm$     0.10 & 2 &    0.230 $\pm$    0.046 &   34.265 $\pm$    1.508 &    1.597/23\\
&  & 10-20\% & 4 &    0.211 $\pm$    0.004 &     0.88 $\pm$     0.08 & 2 &    0.228 $\pm$    0.045 &   25.603 $\pm$    1.203 &    1.361/23\\
&  & 20-30\% & 4 &    0.205 $\pm$    0.004 &     0.85 $\pm$     0.12 & 2 &    0.250 $\pm$    0.050 &   17.849 $\pm$    0.696 &    6.665/23\\
Figure 2(c)&  $p$ & 30-40\% & 4 &    0.200 $\pm$    0.004 &     0.80 $\pm$     0.16 & 2 &    0.285 $\pm$    0.038 &   11.424 $\pm$    0.434 &    5.483/23\\
&  & 40-50\% & 4 &    0.189 $\pm$    0.003 &     0.80 $\pm$     0.16 & 2 &    0.275 $\pm$    0.029 &    7.038 $\pm$    0.282 &   11.643/22\\
&  & 50-60\% & 4 &    0.179 $\pm$    0.003 &     0.70 $\pm$     0.14 & 2 &    0.286 $\pm$    0.017 &    4.076 $\pm$    0.167 &   12.886/22\\
&  & 60-70\% & 4 &    0.165 $\pm$    0.003 &     0.72 $\pm$     0.14 & 2 &    0.270 $\pm$    0.010 &    2.208 $\pm$    0.097 &   57.248/22\\
&  & 70-80\% & 4 &    0.161 $\pm$    0.003 &     0.69 $\pm$     0.13 & 2 &    0.246 $\pm$    0.015 &    0.995 $\pm$    0.040 &   24.989/23\\
\hline
&  & 0-5\% & 4 &    0.216 $\pm$    0.003 &     0.88 $\pm$     0.08 & 2 &    0.234 $\pm$    0.046 &    1.374 $\pm$    0.066 &   23.483/17\\
&  & 5-10\% & 4 &    0.209 $\pm$    0.005 &     0.84 $\pm$     0.14 & 2 &    0.273 $\pm$    0.054 &    1.098 $\pm$    0.047 &   21.419/17\\
&  & 10-20\% & 4 &    0.203 $\pm$    0.004 &     0.85 $\pm$     0.10 & 2 &    0.253 $\pm$    0.050 &    0.888 $\pm$    0.039 &   24.352/17\\
&  & 20-30\% & 4 &    0.196 $\pm$    0.004 &     0.83 $\pm$     0.13 & 2 &    0.265 $\pm$    0.045 &    0.687 $\pm$    0.030 &    8.201/17\\
Figure 2(f)&  $\overline{p}$ & 30-40\% & 4 &    0.189 $\pm$    0.004 &     0.86 $\pm$     0.09 & 2 &    0.229 $\pm$    0.045 &    0.491 $\pm$    0.022 &    4.604/17\\
&  & 40-50\% & 4 &    0.176 $\pm$    0.003 &     0.87 $\pm$     0.09 & 2 &    0.214 $\pm$    0.042 &    0.331 $\pm$    0.016 &    5.954/14\\
&  & 50-60\% & 4 &    0.172 $\pm$    0.004 &     0.85 $\pm$     0.11 & 2 &    0.228 $\pm$    0.045 &    0.213 $\pm$    0.010 &    3.503/13\\
&  & 60-70\% & 4 &    0.155 $\pm$    0.004 &     0.77 $\pm$     0.15 & 2 &    0.261 $\pm$    0.034 &    0.127 $\pm$    0.005 &    3.407/8\\
&  & 70-80\% & 4 &    0.147 $\pm$    0.008 &     0.74 $\pm$     0.14 & 2 &    0.256 $\pm$    0.051 &    0.069 $\pm$    0.003 &    7.499/8\\
\hline
\hline
\end{tabular}
\end{center}
}} }

\newpage

{\scriptsize {Table 3. Values of free parameters, normalization constant, and $\chi^2$/dof  corresponding to the two-component Erlang $p_T$ distribution for productions in Au+Au collisions at $\sqrt{s_{NN}}=\rm{27\, GeV}$ for different centralities in Figure 3.
{%
\begin{center}
\begin{tabular}{ccccccccccc}
\hline
\hline
Figure& Particle& Centrality & $m_1$ &$<p_{ti1}>$ & $k_1$ & $m_2$ &$<p_{ti2}>$ & $N_0$& $\chi^2$/dof \\
 &  (GeV) &  &   & (GeV/c) &  & & (GeV/c)& & \\
\hline
&  & 0-5\% & 2 &    0.156 $\pm$    0.008 &     0.57 $\pm$     0.05 & 2 &    0.249 $\pm$    0.005 &  165.077 $\pm$    8.089 &    0.737/20\\
&  & 5-10\% & 2 &    0.155 $\pm$    0.007 &     0.61 $\pm$     0.04 & 2 &    0.253 $\pm$    0.005 &  136.482 $\pm$    5.732 &    0.519/20\\
&  & 10-20\% & 2 &    0.158 $\pm$    0.006 &     0.63 $\pm$     0.04 & 2 &    0.255 $\pm$    0.004 &  103.435 $\pm$    4.137 &    0.375/20\\
&  & 20-30\% & 2 &    0.157 $\pm$    0.006 &     0.66 $\pm$     0.03 & 2 &    0.257 $\pm$    0.005 &   70.526 $\pm$    2.962 &    0.431/20\\
Figure 3(a)&  $\pi^{+}$ & 30-40\% & 2 &    0.150 $\pm$    0.005 &     0.64 $\pm$     0.04 & 2 &    0.250 $\pm$    0.004 &   48.064 $\pm$    1.730 &    0.625/20\\
&  & 40-50\% & 2 &    0.145 $\pm$    0.005 &     0.63 $\pm$     0.04 & 2 &    0.244 $\pm$    0.004 &   30.629 $\pm$    1.072 &    1.211/20\\
&  & 50-60\% & 2 &    0.144 $\pm$    0.005 &     0.65 $\pm$     0.03 & 2 &    0.241 $\pm$    0.004 &   18.732 $\pm$    0.693 &    1.247/20\\
&  & 60-70\% & 2 &    0.147 $\pm$    0.005 &     0.68 $\pm$     0.03 & 2 &    0.238 $\pm$    0.004 &   10.253 $\pm$    0.390 &    0.759/20\\
&  & 70-80\% & 2 &    0.131 $\pm$    0.006 &     0.53 $\pm$     0.03 & 2 &    0.214 $\pm$    0.003 &    5.523 $\pm$    0.232 &    2.812/20\\
\hline
&  & 0-5\% & 2 &    0.145 $\pm$    0.008 &     0.56 $\pm$     0.04 & 2 &    0.246 $\pm$    0.004 &  176.077 $\pm$    8.452 &    0.682/20\\
&  & 5-10\% & 2 &    0.144 $\pm$    0.007 &     0.58 $\pm$     0.04 & 2 &    0.247 $\pm$    0.004 &  145.211 $\pm$    6.244 &    0.511/20\\
&  & 10-20\% & 2 &    0.150 $\pm$    0.007 &     0.61 $\pm$     0.04 & 2 &    0.251 $\pm$    0.005 &  108.850 $\pm$    5.007 &    0.342/20\\
&  & 20-30\% & 2 &    0.147 $\pm$    0.005 &     0.62 $\pm$     0.04 & 2 &    0.250 $\pm$    0.005 &   74.545 $\pm$    2.684 &    1.105/20\\
Figure 3(d)&  $\pi^{-}$ & 30-40\% & 2 &    0.145 $\pm$    0.005 &     0.63 $\pm$     0.04 & 2 &    0.248 $\pm$    0.005 &   50.278 $\pm$    1.860 &    0.946/20\\
&  & 40-50\% & 2 &    0.143 $\pm$    0.005 &     0.63 $\pm$     0.03 & 2 &    0.243 $\pm$    0.003 &   31.803 $\pm$    1.177 &    1.114/20\\
&  & 50-60\% & 2 &    0.139 $\pm$    0.005 &     0.61 $\pm$     0.03 & 2 &    0.235 $\pm$    0.003 &   19.461 $\pm$    0.681 &    0.970/20\\
&  & 60-70\% & 2 &    0.133 $\pm$    0.006 &     0.56 $\pm$     0.03 & 2 &    0.223 $\pm$    0.003 &   10.814 $\pm$    0.411 &    2.663/20\\
&  & 70-80\% & 2 &    0.129 $\pm$    0.005 &     0.51 $\pm$     0.03 & 2 &    0.211 $\pm$    0.002 &    5.709 $\pm$    0.183 &    3.269/19\\
\hline
&  & 0-5\% & 3 &    0.198 $\pm$    0.005 &     0.67 $\pm$     0.13 & 2 &    0.294 $\pm$    0.009 &   29.706 $\pm$    0.891 &    1.316/20\\
&  & 5-10\% & 3 &    0.192 $\pm$    0.005 &     0.68 $\pm$     0.07 & 2 &    0.315 $\pm$    0.007 &   24.335 $\pm$    0.973 &    1.187/20\\
&  & 10-20\% & 3 &    0.192 $\pm$    0.006 &     0.57 $\pm$     0.10 & 2 &    0.291 $\pm$    0.008 &   18.218 $\pm$    0.619 &    1.652/20\\
&  & 20-30\% & 3 &    0.186 $\pm$    0.006 &     0.53 $\pm$     0.09 & 2 &    0.285 $\pm$    0.005 &   12.546 $\pm$    0.452 &    1.466/20\\
Figure 3(b)& $K^{+}$ & 30-40\% & 3 &    0.198 $\pm$    0.002 &     0.69 $\pm$     0.13 & 2 &    0.223 $\pm$    0.009 &    8.233 $\pm$    0.255 &   26.212/20\\
&  & 40-50\% & 3 &    0.194 $\pm$    0.002 &     0.63 $\pm$     0.12 & 2 &    0.220 $\pm$    0.007 &    4.992 $\pm$    0.155 &   13.256/19\\
&  & 50-60\% & 3 &    0.165 $\pm$    0.005 &     0.51 $\pm$     0.05 & 2 &    0.266 $\pm$    0.004 &    2.790 $\pm$    0.089 &    3.925/19\\
&  & 60-70\% & 3 &    0.152 $\pm$    0.005 &     0.51 $\pm$     0.04 & 2 &    0.269 $\pm$    0.005 &    1.417 $\pm$    0.054 &    1.169/17\\
&  & 70-80\% & 3 &    0.148 $\pm$    0.004 &     0.57 $\pm$     0.05 & 2 &    0.274 $\pm$    0.006 &    0.680 $\pm$    0.023 &    2.210/16\\
\hline
&  & 0-5\% & 3 &    0.192 $\pm$    0.004 &     0.64 $\pm$     0.07 & 2 &    0.292 $\pm$    0.007 &   18.620 $\pm$    0.596 &    2.950/20\\
&  & 5-10\% & 3 &    0.193 $\pm$    0.003 &     0.70 $\pm$     0.09 & 2 &    0.284 $\pm$    0.006 &   15.498 $\pm$    0.527 &    1.836/20\\
&  & 10-20\% & 3 &    0.199 $\pm$    0.003 &     0.65 $\pm$     0.13 & 2 &    0.248 $\pm$    0.008 &   11.714 $\pm$    0.375 &    1.861/20\\
&  & 20-30\% & 3 &    0.183 $\pm$    0.005 &     0.56 $\pm$     0.08 & 2 &    0.274 $\pm$    0.005 &    8.148 $\pm$    0.253 &    2.810/20\\
Figure 3(e)& $K^{-}$ & 30-40\% & 3 &    0.174 $\pm$    0.005 &     0.51 $\pm$     0.09 & 2 &    0.270 $\pm$    0.004 &    5.358 $\pm$    0.166 &    2.310/20\\
&  & 40-50\% & 3 &    0.170 $\pm$    0.004 &     0.58 $\pm$     0.06 & 2 &    0.262 $\pm$    0.004 &    3.299 $\pm$    0.102 &    4.047/19\\
&  & 50-60\% & 3 &    0.170 $\pm$    0.004 &     0.51 $\pm$     0.10 & 2 &    0.240 $\pm$    0.005 &    1.920 $\pm$    0.060 &    8.328/19\\
&  & 60-70\% & 3 &    0.172 $\pm$    0.003 &     0.51 $\pm$     0.10 & 2 &    0.216 $\pm$    0.006 &    0.991 $\pm$    0.035 &    8.495/17\\
&  & 70-80\% & 3 &    0.172 $\pm$    0.003 &     0.65 $\pm$     0.05 & 2 &    0.178 $\pm$    0.011 &    0.472 $\pm$    0.015 &   10.403/15\\
\hline
&  & 0-5\% & 4 &    0.222 $\pm$    0.005 &     0.79 $\pm$     0.11 & 2 &    0.278 $\pm$    0.055 &   34.690 $\pm$    1.353 &    6.636/23\\
&  & 5-10\% & 4 &    0.221 $\pm$    0.004 &     0.84 $\pm$     0.08 & 2 &    0.261 $\pm$    0.052 &   28.720 $\pm$    1.120 &    3.465/19\\
&  & 10-20\% & 4 &    0.220 $\pm$    0.003 &     0.85 $\pm$     0.16 & 2 &    0.261 $\pm$    0.052 &   22.471 $\pm$    0.921 &   13.856/17\\
&  & 20-30\% & 4 &    0.208 $\pm$    0.003 &     0.84 $\pm$     0.07 & 2 &    0.250 $\pm$    0.050 &   14.238 $\pm$    0.541 &    3.297/17\\
Figure 3(c)&  $p$ & 30-40\% & 4 &    0.202 $\pm$    0.003 &     0.81 $\pm$     0.10 & 2 &    0.250 $\pm$    0.038 &    9.105 $\pm$    0.355 &    4.214/17\\
&  & 40-50\% & 4 &    0.201 $\pm$    0.003 &     0.82 $\pm$     0.14 & 2 &    0.229 $\pm$    0.044 &    5.738 $\pm$    0.258 &   11.385/17\\
&  & 50-60\% & 4 &    0.191 $\pm$    0.003 &     0.81 $\pm$     0.16 & 2 &    0.212 $\pm$    0.042 &    3.252 $\pm$    0.140 &   17.833/17\\
&  & 60-70\% & 4 &    0.185 $\pm$    0.003 &     0.82 $\pm$     0.09 & 2 &    0.200 $\pm$    0.040 &    1.684 $\pm$    0.082 &   17.194/17\\
&  & 70-80\% & 4 &    0.175 $\pm$    0.003 &     0.78 $\pm$     0.09 & 2 &    0.200 $\pm$    0.032 &    0.818 $\pm$    0.038 &   18.275/17\\
\hline
&  & 0-5\% & 4 &    0.222 $\pm$    0.004 &     0.91 $\pm$     0.08 & 2 &    0.247 $\pm$    0.049 &    3.937 $\pm$    0.161 &   11.586/16\\
&  & 5-10\% & 4 &    0.215 $\pm$    0.003 &     0.88 $\pm$     0.11 & 2 &    0.290 $\pm$    0.058 &    3.204 $\pm$    0.131 &   15.482/16\\
&  & 10-20\% & 4 &    0.213 $\pm$    0.004 &     0.89 $\pm$     0.10 & 2 &    0.240 $\pm$    0.048 &    2.562 $\pm$    0.120 &   12.446/18\\
&  & 20-30\% & 4 &    0.209 $\pm$    0.004 &     0.88 $\pm$     0.11 & 2 &    0.240 $\pm$    0.048 &    1.926 $\pm$    0.089 &    2.372/18\\
Figure 3(f)&  $\overline{p}$ & 30-40\% & 4 &    0.199 $\pm$    0.003 &     0.88 $\pm$     0.08 & 2 &    0.249 $\pm$    0.049 &    1.355 $\pm$    0.066 &    1.118/19\\
&  & 40-50\% & 4 &    0.190 $\pm$    0.004 &     0.78 $\pm$     0.15 & 2 &    0.313 $\pm$    0.029 &    0.940 $\pm$    0.038 &    3.909/19\\
&  & 50-60\% & 4 &    0.177 $\pm$    0.003 &     0.79 $\pm$     0.15 & 2 &    0.300 $\pm$    0.018 &    0.586 $\pm$    0.028 &    1.950/19\\
&  & 60-70\% & 4 &    0.167 $\pm$    0.003 &     0.81 $\pm$     0.15 & 2 &    0.268 $\pm$    0.023 &    0.340 $\pm$    0.014 &    7.299/17\\
&  & 70-80\% & 4 &    0.155 $\pm$    0.003 &     0.81 $\pm$     0.14 & 2 &    0.271 $\pm$    0.033 &    0.169 $\pm$    0.007 &    4.949/16\\
\hline
\hline
\end{tabular}
\end{center}
}} }
\newpage

{\scriptsize {Table 4. Values of free parameters, normalization constant, and $\chi^2$/dof  corresponding to the two-component Erlang $p_T$ distribution for productions in Au+Au collisions at $\sqrt{s_{NN}}=\rm{27\, GeV}$ for different centralities in Figure 4.
{%
\begin{center}
\begin{tabular}{ccccccccccc}
\hline
\hline
Figure& Particle& Centrality & $m_1$ &$<p_{ti1}>$ & $k_1$ & $m_2$ &$<p_{ti2}>$ & $N_0$& $\chi^2$/dof \\
 &  (GeV) &  &   & (GeV/c) &  & & (GeV/c)& & \\
\hline
&  & 0-5\% & 2 &    0.152 $\pm$    0.007 &     0.51 $\pm$     0.05 & 2 &    0.249 $\pm$    0.003 &  182.402 $\pm$    6.202 &    2.449/20\\
&  & 5-10\% & 2 &    0.155 $\pm$    0.007 &     0.60 $\pm$     0.04 & 2 &    0.257 $\pm$    0.005 &  153.577 $\pm$    6.911 &    0.311/20\\
&  & 10-20\% & 2 &    0.161 $\pm$    0.007 &     0.64 $\pm$     0.04 & 2 &    0.263 $\pm$    0.005 &  114.051 $\pm$    4.676 &    0.277/20\\
&  & 20-30\% & 2 &    0.158 $\pm$    0.007 &     0.65 $\pm$     0.04 & 2 &    0.263 $\pm$    0.005 &   78.449 $\pm$    3.060 &    0.423/20\\
Figure 4(a)&  $\pi^{+}$ & 30-40\% & 2 &    0.155 $\pm$    0.005 &     0.66 $\pm$     0.04 & 2 &    0.261 $\pm$    0.005 &   52.828 $\pm$    2.007 &    0.615/20\\
&  & 40-50\% & 2 &    0.160 $\pm$    0.005 &     0.72 $\pm$     0.03 & 2 &    0.268 $\pm$    0.006 &   32.638 $\pm$    1.338 &    0.568/20\\
&  & 50-60\% & 2 &    0.161 $\pm$    0.005 &     0.74 $\pm$     0.03 & 2 &    0.267 $\pm$    0.005 &   19.253 $\pm$    0.712 &    0.489/20\\
&  & 60-70\% & 2 &    0.152 $\pm$    0.004 &     0.71 $\pm$     0.03 & 2 &    0.255 $\pm$    0.004 &   11.041 $\pm$    0.353 &    0.909/20\\
&  & 70-80\% & 2 &    0.160 $\pm$    0.004 &     0.79 $\pm$     0.02 & 2 &    0.264 $\pm$    0.004 &    5.287 $\pm$    0.169 &    1.377/20\\
\hline
&  & 0-5\% & 2 &    0.164 $\pm$    0.006 &     0.66 $\pm$     0.04 & 2 &    0.264 $\pm$    0.004 &  186.402 $\pm$    6.710 &    2.325/20\\
&  & 5-10\% & 2 &    0.149 $\pm$    0.007 &     0.60 $\pm$     0.04 & 2 &    0.255 $\pm$    0.005 &  160.852 $\pm$    6.273 &    0.452/20\\
&  & 10-20\% & 2 &    0.157 $\pm$    0.007 &     0.63 $\pm$     0.04 & 2 &    0.261 $\pm$    0.005 &  117.063 $\pm$    4.800 &    0.337/20\\
&  & 20-30\% & 2 &    0.154 $\pm$    0.006 &     0.63 $\pm$     0.04 & 2 &    0.259 $\pm$    0.005 &   80.615 $\pm$    2.983 &    0.418/20\\
Figure 4(d)&  $\pi^{-}$ & 30-40\% & 2 &    0.160 $\pm$    0.005 &     0.69 $\pm$     0.03 & 2 &    0.267 $\pm$    0.006 &   52.364 $\pm$    1.676 &    0.442/20\\
&  & 40-50\% & 2 &    0.155 $\pm$    0.006 &     0.68 $\pm$     0.03 & 2 &    0.261 $\pm$    0.005 &   33.302 $\pm$    1.565 &    0.349/20\\
&  & 50-60\% & 2 &    0.150 $\pm$    0.005 &     0.69 $\pm$     0.03 & 2 &    0.257 $\pm$    0.005 &   20.431 $\pm$    0.797 &    0.675/20\\
&  & 60-70\% & 2 &    0.152 $\pm$    0.006 &     0.71 $\pm$     0.03 & 2 &    0.254 $\pm$    0.005 &   11.201 $\pm$    0.538 &    0.358/20\\
&  & 70-80\% & 2 &    0.144 $\pm$    0.005 &     0.64 $\pm$     0.03 & 2 &    0.235 $\pm$    0.003 &    5.647 $\pm$    0.209 &    1.634/20\\
\hline
&  & 0-5\% & 3 &    0.205 $\pm$    0.004 &     0.97 $\pm$     0.02 & 2 &    0.575 $\pm$    0.115 &   29.993 $\pm$    0.930 &    1.988/20\\
&  & 5-10\% & 3 &    0.201 $\pm$    0.003 &     0.94 $\pm$     0.02 & 2 &    0.462 $\pm$    0.030 &   24.959 $\pm$    0.799 &    2.304/20\\
&  & 10-20\% & 3 &    0.199 $\pm$    0.003 &     0.92 $\pm$     0.03 & 2 &    0.430 $\pm$    0.028 &   18.851 $\pm$    0.547 &    2.702/20\\
&  & 20-30\% & 3 &    0.191 $\pm$    0.004 &     0.77 $\pm$     0.05 & 2 &    0.349 $\pm$    0.010 &   12.830 $\pm$    0.398 &    1.806/20\\
Figure 4(b)& $K^{+}$ & 30-40\% & 3 &    0.186 $\pm$    0.004 &     0.77 $\pm$     0.03 & 2 &    0.348 $\pm$    0.006 &    8.241 $\pm$    0.247 &    2.330/20\\
&  & 40-50\% & 3 &    0.174 $\pm$    0.004 &     0.60 $\pm$     0.05 & 2 &    0.307 $\pm$    0.006 &    5.274 $\pm$    0.185 &    1.779/20\\
&  & 50-60\% & 3 &    0.171 $\pm$    0.005 &     0.64 $\pm$     0.05 & 2 &    0.305 $\pm$    0.006 &    2.935 $\pm$    0.091 &    2.899/20\\
&  & 60-70\% & 3 &    0.165 $\pm$    0.003 &     0.72 $\pm$     0.03 & 2 &    0.313 $\pm$    0.007 &    1.512 $\pm$    0.045 &    4.541/20\\
&  & 70-80\% & 3 &    0.152 $\pm$    0.005 &     0.51 $\pm$     0.05 & 2 &    0.276 $\pm$    0.004 &    0.702 $\pm$    0.022 &    4.851/20\\
\hline
&  & 0-5\% & 3 &    0.198 $\pm$    0.002 &     0.97 $\pm$     0.01 & 2 &    0.531 $\pm$    0.052 &   21.872 $\pm$    0.634 &    5.463/19\\
&  & 5-10\% & 3 &    0.192 $\pm$    0.003 &     0.85 $\pm$     0.03 & 2 &    0.357 $\pm$    0.012 &   18.306 $\pm$    0.567 &    2.210/20\\
&  & 10-20\% & 3 &    0.188 $\pm$    0.003 &     0.82 $\pm$     0.04 & 2 &    0.351 $\pm$    0.010 &   14.220 $\pm$    0.427 &    2.037/20\\
&  & 20-30\% & 3 &    0.185 $\pm$    0.003 &     0.81 $\pm$     0.03 & 2 &    0.347 $\pm$    0.008 &    9.713 $\pm$    0.291 &    2.793/20\\
Figure 4(e)& $K^{-}$ & 30-40\% & 3 &    0.183 $\pm$    0.003 &     0.83 $\pm$     0.03 & 2 &    0.353 $\pm$    0.009 &    6.140 $\pm$    0.196 &    3.571/20\\
&  & 40-50\% & 3 &    0.172 $\pm$    0.004 &     0.65 $\pm$     0.05 & 2 &    0.302 $\pm$    0.005 &    3.911 $\pm$    0.121 &    2.514/20\\
&  & 50-60\% & 3 &    0.169 $\pm$    0.004 &     0.70 $\pm$     0.04 & 2 &    0.303 $\pm$    0.005 &    2.185 $\pm$    0.076 &    2.350/20\\
&  & 60-70\% & 3 &    0.166 $\pm$    0.003 &     0.82 $\pm$     0.03 & 2 &    0.329 $\pm$    0.009 &    1.101 $\pm$    0.034 &    4.562/20\\
&  & 70-80\% & 3 &    0.163 $\pm$    0.004 &     0.72 $\pm$     0.05 & 2 &    0.276 $\pm$    0.008 &    0.512 $\pm$    0.016 &   12.189/20\\
\hline
&  & 0-5\% & 4 &    0.226 $\pm$    0.005 &     0.80 $\pm$     0.16 & 2 &    0.284 $\pm$    0.056 &   30.191 $\pm$    1.328 &    8.619/17\\
&  & 5-10\% & 4 &    0.222 $\pm$    0.005 &     0.82 $\pm$     0.10 & 2 &    0.261 $\pm$    0.052 &   26.024 $\pm$    1.093 &    7.610/17\\
&  & 10-20\% & 4 &    0.222 $\pm$    0.004 &     0.83 $\pm$     0.10 & 2 &    0.261 $\pm$    0.052 &   20.160 $\pm$    0.887 &    4.736/17\\
&  & 20-30\% & 4 &    0.216 $\pm$    0.004 &     0.78 $\pm$     0.15 & 2 &    0.308 $\pm$    0.035 &   13.750 $\pm$    0.577 &    9.822/17\\
Figure 4(c)&  $p$ & 30-40\% & 4 &    0.206 $\pm$    0.004 &     0.70 $\pm$     0.14 & 2 &    0.345 $\pm$    0.022 &    9.066 $\pm$    0.381 &   13.078/17\\
&  & 40-50\% & 4 &    0.196 $\pm$    0.006 &     0.65 $\pm$     0.13 & 2 &    0.339 $\pm$    0.020 &    5.287 $\pm$    0.248 &    2.337/17\\
&  & 50-60\% & 4 &    0.185 $\pm$    0.004 &     0.67 $\pm$     0.13 & 2 &    0.318 $\pm$    0.018 &    3.037 $\pm$    0.118 &    6.823/17\\
&  & 60-70\% & 4 &    0.175 $\pm$    0.004 &     0.64 $\pm$     0.12 & 2 &    0.305 $\pm$    0.013 &    1.725 $\pm$    0.069 &   23.466/17\\
&  & 70-80\% & 4 &    0.160 $\pm$    0.004 &     0.57 $\pm$     0.08 & 2 &    0.293 $\pm$    0.008 &    0.733 $\pm$    0.029 &    5.857/17\\
\hline
&  & 0-5\% & 4 &    0.228 $\pm$    0.006 &     0.78 $\pm$     0.15 & 2 &    0.343 $\pm$    0.059 &    5.877 $\pm$    0.235 &   12.404/16\\
&  & 5-10\% & 4 &    0.228 $\pm$    0.005 &     0.86 $\pm$     0.10 & 2 &    0.260 $\pm$    0.052 &    4.869 $\pm$    0.234 &    9.255/16\\
&  & 10-20\% & 4 &    0.222 $\pm$    0.004 &     0.87 $\pm$     0.12 & 2 &    0.240 $\pm$    0.048 &    3.884 $\pm$    0.179 &    8.533/16\\
&  & 20-30\% & 4 &    0.217 $\pm$    0.004 &     0.85 $\pm$     0.14 & 2 &    0.241 $\pm$    0.048 &    2.900 $\pm$    0.122 &    1.710/16\\
Figure 4(f)&  $\overline{p}$ & 30-40\% & 4 &    0.210 $\pm$    0.003 &     0.83 $\pm$     0.13 & 2 &    0.249 $\pm$    0.047 &    2.072 $\pm$    0.085 &    1.695/16\\
&  & 40-50\% & 4 &    0.200 $\pm$    0.004 &     0.74 $\pm$     0.14 & 2 &    0.295 $\pm$    0.025 &    1.399 $\pm$    0.062 &    1.579/16\\
&  & 50-60\% & 4 &    0.185 $\pm$    0.004 &     0.73 $\pm$     0.14 & 2 &    0.300 $\pm$    0.017 &    0.820 $\pm$    0.032 &    2.232/16\\
&  & 60-70\% & 4 &    0.177 $\pm$    0.003 &     0.76 $\pm$     0.15 & 2 &    0.277 $\pm$    0.017 &    0.482 $\pm$    0.021 &   21.340/16\\
&  & 70-80\% & 4 &    0.157 $\pm$    0.003 &     0.67 $\pm$     0.13 & 2 &    0.278 $\pm$    0.012 &    0.229 $\pm$    0.009 &    4.569/14\\
\hline
\hline
\end{tabular}
\end{center}
}} }

\newpage

{\scriptsize {Table 5. Values of free parameters, normalization constant, and $\chi^2$/dof  corresponding to the two-component Erlang $p_T$ distribution for productions in Au+Au collisions at $\sqrt{s_{NN}}=\rm{39\, GeV}$ for different centralities in Figure 5.
{%
\begin{center}
\begin{tabular}{ccccccccccc}
\hline
\hline
Figure& Particle& Centrality & $m_1$ &$<p_{ti1}>$ & $k_1$ & $m_2$ &$<p_{ti2}>$ & $N_0$& $\chi^2$/dof \\
 &  (GeV) &  &   & (GeV/c) &  & & (GeV/c)& & \\
\hline
&  & 0-5\% & 2 &    0.155 $\pm$    0.008 &     0.54 $\pm$     0.05 & 2 &    0.265 $\pm$    0.004 &  185.159 $\pm$    7.406 &    3.265/20\\
&  & 5-10\% & 2 &    0.166 $\pm$    0.007 &     0.63 $\pm$     0.04 & 2 &    0.274 $\pm$    0.006 &  153.984 $\pm$    6.159 &    0.386/20\\
&  & 10-20\% & 2 &    0.158 $\pm$    0.008 &     0.61 $\pm$     0.04 & 2 &    0.270 $\pm$    0.005 &  121.765 $\pm$    6.332 &    0.376/20\\
&  & 20-30\% & 2 &    0.157 $\pm$    0.008 &     0.63 $\pm$     0.04 & 2 &    0.272 $\pm$    0.005 &   83.486 $\pm$    4.091 &    0.484/20\\
Figure 5(a)&  $\pi^{+}$ & 30-40\% & 2 &    0.158 $\pm$    0.007 &     0.66 $\pm$     0.04 & 2 &    0.275 $\pm$    0.006 &   54.946 $\pm$    2.692 &    0.493/20\\
&  & 40-50\% & 2 &    0.160 $\pm$    0.006 &     0.70 $\pm$     0.03 & 2 &    0.279 $\pm$    0.006 &   34.963 $\pm$    1.538 &    0.577/20\\
&  & 50-60\% & 2 &    0.156 $\pm$    0.004 &     0.72 $\pm$     0.03 & 2 &    0.279 $\pm$    0.006 &   21.974 $\pm$    0.769 &    0.740/20\\
&  & 60-70\% & 2 &    0.154 $\pm$    0.005 &     0.72 $\pm$     0.03 & 2 &    0.273 $\pm$    0.005 &   12.108 $\pm$    0.533 &    0.446/20\\
&  & 70-80\% & 2 &    0.153 $\pm$    0.006 &     0.73 $\pm$     0.03 & 2 &    0.272 $\pm$    0.006 &    6.668 $\pm$    0.287 &    0.419/20\\
\hline
&  & 0-5\% & 2 &    0.153 $\pm$    0.010 &     0.53 $\pm$     0.04 & 2 &    0.258 $\pm$    0.004 &  191.409 $\pm$    7.274 &    1.089/20\\
&  & 5-10\% & 2 &    0.159 $\pm$    0.008 &     0.60 $\pm$     0.04 & 2 &    0.266 $\pm$    0.006 &  159.491 $\pm$    7.975 &    0.233/20\\
&  & 10-20\% & 2 &    0.152 $\pm$    0.008 &     0.59 $\pm$     0.04 & 2 &    0.264 $\pm$    0.005 &  126.386 $\pm$    6.446 &    0.306/20\\
&  & 20-30\% & 2 &    0.155 $\pm$    0.007 &     0.63 $\pm$     0.04 & 2 &    0.269 $\pm$    0.005 &   85.965 $\pm$    4.298 &    0.365/20\\
Figure 5(d)&  $\pi^{-}$ & 30-40\% & 2 &    0.156 $\pm$    0.007 &     0.65 $\pm$     0.04 & 2 &    0.271 $\pm$    0.006 &   56.262 $\pm$    2.869 &    0.355/20\\
&  & 40-50\% & 2 &    0.158 $\pm$    0.006 &     0.69 $\pm$     0.03 & 2 &    0.274 $\pm$    0.006 &   35.958 $\pm$    1.654 &    0.376/20\\
&  & 50-60\% & 2 &    0.155 $\pm$    0.006 &     0.71 $\pm$     0.03 & 2 &    0.273 $\pm$    0.006 &   22.498 $\pm$    1.080 &    0.483/20\\
&  & 60-70\% & 2 &    0.156 $\pm$    0.006 &     0.73 $\pm$     0.03 & 2 &    0.273 $\pm$    0.005 &   12.258 $\pm$    0.576 &    0.425/20\\
&  & 70-80\% & 2 &    0.153 $\pm$    0.006 &     0.72 $\pm$     0.03 & 2 &    0.266 $\pm$    0.006 &    6.781 $\pm$    0.353 &    0.341/20\\
\hline
&  & 0-5\% & 3 &    0.211 $\pm$    0.003 &     0.94 $\pm$     0.06 & 2 &    0.359 $\pm$    0.045 &   31.219 $\pm$    1.030 &    5.650/20\\
&  & 5-10\% & 3 &    0.202 $\pm$    0.004 &     0.83 $\pm$     0.05 & 2 &    0.369 $\pm$    0.014 &   27.111 $\pm$    0.840 &    1.564/20\\
&  & 10-20\% & 3 &    0.198 $\pm$    0.005 &     0.73 $\pm$     0.06 & 2 &    0.349 $\pm$    0.010 &   20.074 $\pm$    0.642 &    1.340/20\\
&  & 20-30\% & 3 &    0.191 $\pm$    0.006 &     0.67 $\pm$     0.06 & 2 &    0.345 $\pm$    0.009 &   13.603 $\pm$    0.422 &    2.281/20\\
Figure 5(b)& $K^{+}$ & 30-40\% & 3 &    0.189 $\pm$    0.005 &     0.68 $\pm$     0.05 & 2 &    0.337 $\pm$    0.006 &    8.778 $\pm$    0.255 &    2.462/20\\
&  & 40-50\% & 3 &    0.174 $\pm$    0.006 &     0.60 $\pm$     0.05 & 2 &    0.335 $\pm$    0.008 &    5.514 $\pm$    0.210 &    1.220/20\\
&  & 50-60\% & 3 &    0.168 $\pm$    0.006 &     0.51 $\pm$     0.06 & 2 &    0.304 $\pm$    0.004 &    3.273 $\pm$    0.098 &    2.751/20\\
&  & 60-70\% & 3 &    0.170 $\pm$    0.004 &     0.63 $\pm$     0.04 & 2 &    0.330 $\pm$    0.006 &    1.605 $\pm$    0.048 &    2.878/20\\
&  & 70-80\% & 3 &    0.164 $\pm$    0.005 &     0.53 $\pm$     0.04 & 2 &    0.304 $\pm$    0.005 &    0.841 $\pm$    0.027 &    6.889/20\\
\hline
&  & 0-5\% & 3 &    0.206 $\pm$    0.003 &     0.86 $\pm$     0.06 & 2 &    0.352 $\pm$    0.020 &   24.658 $\pm$    0.715 &    5.446/20\\
&  & 5-10\% & 3 &    0.199 $\pm$    0.004 &     0.83 $\pm$     0.05 & 2 &    0.354 $\pm$    0.014 &   21.186 $\pm$    0.678 &    3.110/20\\
&  & 10-20\% & 3 &    0.195 $\pm$    0.005 &     0.73 $\pm$     0.05 & 2 &    0.339 $\pm$    0.009 &   15.792 $\pm$    0.553 &    2.063/20\\
&  & 20-30\% & 3 &    0.190 $\pm$    0.006 &     0.65 $\pm$     0.07 & 2 &    0.321 $\pm$    0.008 &   10.783 $\pm$    0.356 &    2.454/20\\
Figure 5(e)& $K^{-}$ & 30-40\% & 3 &    0.185 $\pm$    0.005 &     0.68 $\pm$     0.05 & 2 &    0.333 $\pm$    0.006 &    7.005 $\pm$    0.224 &    1.989/20\\
&  & 40-50\% & 3 &    0.176 $\pm$    0.005 &     0.56 $\pm$     0.05 & 2 &    0.301 $\pm$    0.005 &    4.478 $\pm$    0.134 &    2.504/20\\
&  & 50-60\% & 3 &    0.169 $\pm$    0.005 &     0.55 $\pm$     0.05 & 2 &    0.294 $\pm$    0.005 &    2.666 $\pm$    0.080 &    2.100/20\\
&  & 60-70\% & 3 &    0.168 $\pm$    0.004 &     0.66 $\pm$     0.04 & 2 &    0.317 $\pm$    0.005 &    1.316 $\pm$    0.039 &    3.159/20\\
&  & 70-80\% & 3 &    0.166 $\pm$    0.004 &     0.67 $\pm$     0.04 & 2 &    0.307 $\pm$    0.007 &    0.677 $\pm$    0.022 &    6.523/20\\
\hline
&  & 0-5\% & 4 &    0.239 $\pm$    0.005 &     0.79 $\pm$     0.09 & 2 &    0.293 $\pm$    0.055 &   26.115 $\pm$    1.097 &    3.078/16\\
&  & 5-10\% & 4 &    0.229 $\pm$    0.004 &     0.86 $\pm$     0.08 & 2 &    0.240 $\pm$    0.048 &   22.026 $\pm$    1.035 &    4.637/16\\
&  & 10-20\% & 4 &    0.229 $\pm$    0.006 &     0.82 $\pm$     0.10 & 2 &    0.281 $\pm$    0.056 &   17.136 $\pm$    0.788 &    1.181/16\\
&  & 20-30\% & 4 &    0.226 $\pm$    0.006 &     0.80 $\pm$     0.16 & 2 &    0.298 $\pm$    0.059 &   12.027 $\pm$    0.469 &    5.461/16\\
Figure 5(c)&  $p$ & 30-40\% & 4 &    0.209 $\pm$    0.006 &     0.69 $\pm$     0.13 & 2 &    0.355 $\pm$    0.032 &    8.191 $\pm$    0.360 &   10.732/16\\
&  & 40-50\% & 4 &    0.209 $\pm$    0.004 &     0.75 $\pm$     0.15 & 2 &    0.318 $\pm$    0.032 &    4.934 $\pm$    0.227 &   13.569/16\\
&  & 50-60\% & 4 &    0.196 $\pm$    0.005 &     0.67 $\pm$     0.13 & 2 &    0.341 $\pm$    0.026 &    2.839 $\pm$    0.128 &    7.375/16\\
&  & 60-70\% & 4 &    0.178 $\pm$    0.004 &     0.62 $\pm$     0.11 & 2 &    0.342 $\pm$    0.015 &    1.411 $\pm$    0.062 &    2.427/16\\
&  & 70-80\% & 4 &    0.171 $\pm$    0.005 &     0.61 $\pm$     0.10 & 2 &    0.317 $\pm$    0.009 &    0.717 $\pm$    0.030 &    5.298/16\\
\hline
&  & 0-5\% & 4 &    0.237 $\pm$    0.004 &     0.88 $\pm$     0.08 & 2 &    0.270 $\pm$    0.054 &    8.086 $\pm$    0.380 &    9.659/17\\
&  & 5-10\% & 4 &    0.231 $\pm$    0.004 &     0.85 $\pm$     0.11 & 2 &    0.310 $\pm$    0.062 &    6.970 $\pm$    0.293 &    9.818/17\\
&  & 10-20\% & 4 &    0.228 $\pm$    0.005 &     0.82 $\pm$     0.12 & 2 &    0.306 $\pm$    0.060 &    5.318 $\pm$    0.261 &    6.490/17\\
&  & 20-30\% & 4 &    0.218 $\pm$    0.004 &     0.86 $\pm$     0.10 & 2 &    0.255 $\pm$    0.051 &    3.722 $\pm$    0.164 &    8.283/17\\
Figure 5(f)&  $\overline{p}$ & 30-40\% & 4 &    0.213 $\pm$    0.004 &     0.83 $\pm$     0.13 & 2 &    0.274 $\pm$    0.054 &    2.756 $\pm$    0.168 &    0.376/17\\
&  & 40-50\% & 4 &    0.204 $\pm$    0.004 &     0.74 $\pm$     0.14 & 2 &    0.325 $\pm$    0.029 &    1.824 $\pm$    0.071 &    3.573/17\\
&  & 50-60\% & 4 &    0.189 $\pm$    0.003 &     0.72 $\pm$     0.14 & 2 &    0.311 $\pm$    0.013 &    1.152 $\pm$    0.045 &    3.976/17\\
&  & 60-70\% & 4 &    0.182 $\pm$    0.003 &     0.75 $\pm$     0.15 & 2 &    0.296 $\pm$    0.013 &    0.636 $\pm$    0.028 &   10.249/17\\
&  & 70-80\% & 4 &    0.172 $\pm$    0.003 &     0.71 $\pm$     0.14 & 2 &    0.281 $\pm$    0.014 &    0.335 $\pm$    0.014 &   23.783/17\\
\hline
\hline
\end{tabular}
\end{center}
}} }

\newpage

{\scriptsize {Table 6. Values of free parameters, normalization constant, and $\chi^2$/dof  corresponding to the two-component Erlang $p_T$ distribution for productions in Au+Au collisions at $\sqrt{s_{NN}}=\rm{62.4\, GeV}$ for different centralities in Figure 6.
{%
\begin{center}
\begin{tabular}{ccccccccccc}
\hline
\hline
Figure& Particle& Centrality & $m_1$ &$<p_{ti1}>$ & $k_1$ & $m_2$ &$<p_{ti2}>$ & $N_0$& $\chi^2$/dof \\
 &  (GeV) &  &   & (GeV/c) &  & & (GeV/c)& & \\
\hline
&  & 0-5\% & 2 &    0.172 $\pm$    0.003 &     0.65 $\pm$     0.05 & 2 &    0.274 $\pm$    0.012 &  232.461 $\pm$    1.860 &    0.261/4\\
&  & 5-10\% & 2 &    0.188 $\pm$    0.003 &     0.85 $\pm$     0.04 & 2 &    0.314 $\pm$    0.022 &  187.816 $\pm$    1.315 &    0.361/4\\
&  & 10-20\% & 2 &    0.156 $\pm$    0.003 &     0.51 $\pm$     0.03 & 2 &    0.261 $\pm$    0.009 &  146.014 $\pm$    1.168 &    0.344/4\\
&  & 20-30\% & 2 &    0.158 $\pm$    0.002 &     0.51 $\pm$     0.03 & 2 &    0.257 $\pm$    0.008 &   99.959 $\pm$    0.800 &    0.468/4\\
Figure 5(a)&  $\pi^{+}$ & 30-40\% & 2 &    0.153 $\pm$    0.003 &     0.51 $\pm$     0.03 & 2 &    0.256 $\pm$    0.009 &   67.869 $\pm$    0.611 &    0.406/4\\
&  & 40-50\% & 2 &    0.147 $\pm$    0.002 &     0.51 $\pm$     0.03 & 2 &    0.254 $\pm$    0.008 &   44.560 $\pm$    0.356 &    0.563/4\\
&  & 50-60\% & 2 &    0.145 $\pm$    0.002 &     0.51 $\pm$     0.02 & 2 &    0.245 $\pm$    0.008 &   27.279 $\pm$    0.164 &    1.851/4\\
&  & 60-70\% & 2 &    0.141 $\pm$    0.002 &     0.53 $\pm$     0.02 & 2 &    0.241 $\pm$    0.011 &   15.409 $\pm$    0.092 &    0.508/4\\
&  & 70-80\% & 2 &    0.136 $\pm$    0.002 &     0.51 $\pm$     0.02 & 2 &    0.230 $\pm$    0.008 &    7.645 $\pm$    0.054 &    1.175/4\\
\hline
&  & 0-5\% & 2 &    0.175 $\pm$    0.002 &     0.67 $\pm$     0.05 & 2 &    0.269 $\pm$    0.008 &  234.954 $\pm$    1.410 &    0.825/4\\
&  & 5-10\% & 2 &    0.179 $\pm$    0.003 &     0.78 $\pm$     0.05 & 2 &    0.278 $\pm$    0.026 &  193.787 $\pm$    2.519 &   24.657/4\\
&  & 10-20\% & 2 &    0.161 $\pm$    0.003 &     0.56 $\pm$     0.04 & 2 &    0.267 $\pm$    0.009 &  147.565 $\pm$    1.328 &    0.072/4\\
&  & 20-30\% & 2 &    0.164 $\pm$    0.003 &     0.59 $\pm$     0.04 & 2 &    0.265 $\pm$    0.011 &  101.980 $\pm$    0.918 &    0.091/4\\
Figure 5(d)&  $\pi^{-}$ & 30-40\% & 2 &    0.159 $\pm$    0.003 &     0.58 $\pm$     0.04 & 2 &    0.269 $\pm$    0.012 &   68.737 $\pm$    0.619 &    0.120/4\\
&  & 40-50\% & 2 &    0.148 $\pm$    0.002 &     0.51 $\pm$     0.04 & 2 &    0.257 $\pm$    0.007 &   44.980 $\pm$    0.270 &    0.867/4\\
&  & 50-60\% & 2 &    0.165 $\pm$    0.002 &     0.75 $\pm$     0.06 & 2 &    0.257 $\pm$    0.018 &   27.748 $\pm$    0.361 &   48.424/4\\
&  & 60-70\% & 2 &    0.166 $\pm$    0.002 &     0.81 $\pm$     0.04 & 2 &    0.263 $\pm$    0.021 &   15.342 $\pm$    0.199 &   78.440/4\\
&  & 70-80\% & 2 &    0.138 $\pm$    0.002 &     0.51 $\pm$     0.02 & 2 &    0.228 $\pm$    0.008 &    7.692 $\pm$    0.046 &    0.572/4\\
\hline
&  & 0-5\% & 3 &    0.246 $\pm$    0.004 &     0.75 $\pm$     0.04 & 2 &    0.256 $\pm$    0.015 &   39.598 $\pm$    0.752 &    0.299/4\\
&  & 5-10\% & 3 &    0.236 $\pm$    0.004 &     0.63 $\pm$     0.10 & 2 &    0.310 $\pm$    0.010 &   32.688 $\pm$    0.490 &    1.425/4\\
&  & 10-20\% & 3 &    0.248 $\pm$    0.003 &     0.69 $\pm$     0.06 & 2 &    0.269 $\pm$    0.009 &   24.648 $\pm$    0.320 &    1.034/4\\
&  & 20-30\% & 3 &    0.226 $\pm$    0.008 &     0.57 $\pm$     0.11 & 2 &    0.302 $\pm$    0.020 &   15.892 $\pm$    0.445 &    5.988/4\\
Figure 5(b)& $K^{+}$ & 30-40\% & 3 &    0.244 $\pm$    0.008 &     0.61 $\pm$     0.10 & 2 &    0.272 $\pm$    0.019 &   11.179 $\pm$    0.291 &    7.769/4\\
&  & 40-50\% & 3 &    0.222 $\pm$    0.005 &     0.57 $\pm$     0.11 & 2 &    0.286 $\pm$    0.013 &    7.046 $\pm$    0.127 &   11.218/4\\
&  & 50-60\% & 3 &    0.234 $\pm$    0.006 &     0.51 $\pm$     0.07 & 2 &    0.275 $\pm$    0.010 &    4.065 $\pm$    0.081 &    0.345/4\\
&  & 60-70\% & 3 &    0.212 $\pm$    0.005 &     0.64 $\pm$     0.12 & 2 &    0.226 $\pm$    0.013 &    2.147 $\pm$    0.045 &    6.846/4\\
&  & 70-80\% & 3 &    0.208 $\pm$    0.011 &     0.61 $\pm$     0.12 & 2 &    0.229 $\pm$    0.020 &    0.934 $\pm$    0.031 &    6.990/4\\
\hline
&  & 0-5\% & 3 &    0.233 $\pm$    0.021 &     0.73 $\pm$     0.14 & 2 &    0.289 $\pm$    0.057 &   33.071 $\pm$    1.885 &   17.735/4\\
&  & 5-10\% & 3 &    0.226 $\pm$    0.004 &     0.73 $\pm$     0.14 & 2 &    0.288 $\pm$    0.017 &   26.889 $\pm$    0.511 &    3.348/4\\
&  & 10-20\% & 3 &    0.228 $\pm$    0.003 &     0.72 $\pm$     0.12 & 2 &    0.291 $\pm$    0.010 &   20.235 $\pm$    0.243 &    3.616/4\\
&  & 20-30\% & 3 &    0.222 $\pm$    0.003 &     0.71 $\pm$     0.14 & 2 &    0.292 $\pm$    0.010 &   14.319 $\pm$    0.172 &    3.191/4\\
Figure 5(e)& $K^{-}$ & 30-40\% & 3 &    0.218 $\pm$    0.004 &     0.71 $\pm$     0.09 & 2 &    0.279 $\pm$    0.015 &    9.119 $\pm$    0.173 &    2.298/4\\
&  & 40-50\% & 3 &    0.208 $\pm$    0.003 &     0.71 $\pm$     0.14 & 2 &    0.265 $\pm$    0.010 &    5.810 $\pm$    0.070 &    5.979/4\\
&  & 50-60\% & 3 &    0.195 $\pm$    0.004 &     0.66 $\pm$     0.13 & 2 &    0.282 $\pm$    0.017 &    3.421 $\pm$    0.072 &    4.352/4\\
&  & 60-70\% & 3 &    0.187 $\pm$    0.003 &     0.71 $\pm$     0.14 & 2 &    0.261 $\pm$    0.012 &    1.777 $\pm$    0.023 &    5.638/4\\
&  & 70-80\% & 3 &    0.183 $\pm$    0.012 &     0.81 $\pm$     0.16 & 2 &    0.203 $\pm$    0.040 &    0.788 $\pm$    0.037 &   13.299/4\\
\hline
&  & 0-5\% & 4 &    0.256 $\pm$    0.002 &     0.92 $\pm$     0.04 & 2 &    0.336 $\pm$    0.031 &   28.857 $\pm$    0.231 &    4.730/9\\
&  & 5-10\% & 4 &    0.253 $\pm$    0.001 &     0.91 $\pm$     0.03 & 2 &    0.382 $\pm$    0.016 &   23.302 $\pm$    0.140 &    7.145/9\\
&  & 10-20\% & 4 &    0.247 $\pm$    0.001 &     0.93 $\pm$     0.04 & 2 &    0.347 $\pm$    0.027 &   17.611 $\pm$    0.123 &    6.489/9\\
&  & 20-30\% & 4 &    0.240 $\pm$    0.002 &     0.94 $\pm$     0.02 & 2 &    0.315 $\pm$    0.045 &   11.620 $\pm$    0.116 &   34.288/9\\
Figure 5(c)&  $p$ & 30-40\% & 4 &    0.231 $\pm$    0.001 &     0.94 $\pm$     0.03 & 2 &    0.283 $\pm$    0.038 &    7.729 $\pm$    0.046 &   13.509/9\\
&  & 40-50\% & 4 &    0.223 $\pm$    0.001 &     0.91 $\pm$     0.04 & 2 &    0.340 $\pm$    0.022 &    4.845 $\pm$    0.029 &   17.310/9\\
&  & 50-60\% & 4 &    0.209 $\pm$    0.001 &     0.89 $\pm$     0.06 & 2 &    0.370 $\pm$    0.029 &    2.854 $\pm$    0.017 &   19.511/9\\
&  & 60-70\% & 4 &    0.195 $\pm$    0.002 &     0.89 $\pm$     0.06 & 2 &    0.370 $\pm$    0.047 &    1.465 $\pm$    0.010 &   32.579/9\\
&  & 70-80\% & 4 &    0.188 $\pm$    0.002 &     0.94 $\pm$     0.04 & 2 &    0.217 $\pm$    0.043 &    0.647 $\pm$    0.006 &   20.348/9\\
\hline
&  & 0-5\% & 4 &    0.289 $\pm$    0.002 &     0.77 $\pm$     0.01 & 2 &    0.395 $\pm$    0.007 &   15.211 $\pm$    0.122 &   10.907/10\\
&  & 5-10\% & 4 &    0.280 $\pm$    0.002 &     0.81 $\pm$     0.03 & 2 &    0.355 $\pm$    0.011 &   12.675 $\pm$    0.139 &   21.127/10\\
&  & 10-20\% & 4 &    0.269 $\pm$    0.002 &     0.84 $\pm$     0.01 & 2 &    0.312 $\pm$    0.013 &    9.551 $\pm$    0.105 &   13.318/10\\
&  & 20-30\% & 4 &    0.255 $\pm$    0.002 &     0.82 $\pm$     0.02 & 2 &    0.312 $\pm$    0.018 &    6.532 $\pm$    0.046 &   13.768/10\\
Figure 5(f)&  $\overline{p}$ & 30-40\% & 4 &    0.239 $\pm$    0.002 &     0.75 $\pm$     0.04 & 2 &    0.357 $\pm$    0.013 &    4.335 $\pm$    0.030 &   15.160/10\\
&  & 40-50\% & 4 &    0.223 $\pm$    0.002 &     0.84 $\pm$     0.02 & 2 &    0.264 $\pm$    0.020 &    2.811 $\pm$    0.034 &   20.758/10\\
&  & 50-60\% & 4 &    0.208 $\pm$    0.002 &     0.79 $\pm$     0.05 & 2 &    0.307 $\pm$    0.011 &    1.688 $\pm$    0.014 &    6.896/10\\
&  & 60-70\% & 4 &    0.198 $\pm$    0.002 &     0.74 $\pm$     0.05 & 2 &    0.320 $\pm$    0.021 &    0.963 $\pm$    0.009 &   23.599/10\\
&  & 70-80\% & 4 &    0.180 $\pm$    0.003 &     0.73 $\pm$     0.06 & 2 &    0.313 $\pm$    0.034 &    0.433 $\pm$    0.005 &   27.562/10\\
\hline
\hline
\end{tabular}
\end{center}
}} }

According to the extracted normalization constants from the above comparisons, the yield ratios of negative to positive particles, $k_{\pi}$, $k_{K}$, and $k_{p}$, versus collision energy and centrality are obtained. The three types of yield ratios show regular trends with increase of collision energy and  centrality. In order to see the dependences of the three yield ratios on centrality, Figures 7--9 respectively show the change trends of the three yield ratios of $k_{\pi}$, $k_{K}$, and $k_{p}$ with different centralities at different energies. As can be seen, $k_{\pi}$ varies around 1.05 and decreases with increase of energy, but do not shows obvious dependence on centrality. $k_{K}$ varies between 0.35 and  0.85, and increases obviously with increase of energy. At some energies (7.7, 11.5, 19.6 and 39 GeV), $k_{K}$ increases with increase of centrality class, but at these energies of 27 and 62.4 GeV, $k_{K}$ do not shows obvious dependence on centrality. $k_{p}$ varies between 0.007 and 0.7, and increases obviously with increase of energy. Unlike $k_{\pi}$ and $k_{K}$, $k_{p}$ increases obviously with increase of centrality class at all energies considered in this paper, which means that $k_{p}$ shows obvious dependence on centrality. Overall, the dependence of $k_{p}$ on centrality is higher than that of $k_{K}$, and the dependence of $k_{K}$ on centrality is higher than that of $k_{\pi}$, which indicates that the correlation between the generation mechanism of $p$ ($\bar{p}$) and centrality is relatively the highest, followed by $K^{\pm}$, and $\pi^{\pm}$ is the weakest. In addition, it is not difficult to notice that with the increase of energy, the values of $k_{K}$ and $k_{p}$ are both less than 1 and gradually increase (most cases), while that of $k_{\pi}$ are almost equal to 1, which indicates that the generation mechanisms of these particles are closely related to the collision energy, and the generation mechanism of $p$ and $\bar{p}$ is similar to $K^{\pm}$, but different from $\pi^{\pm}$.
\begin{figure}[H]
\hskip-0.0cm {\centering
\includegraphics[width=16.0cm]{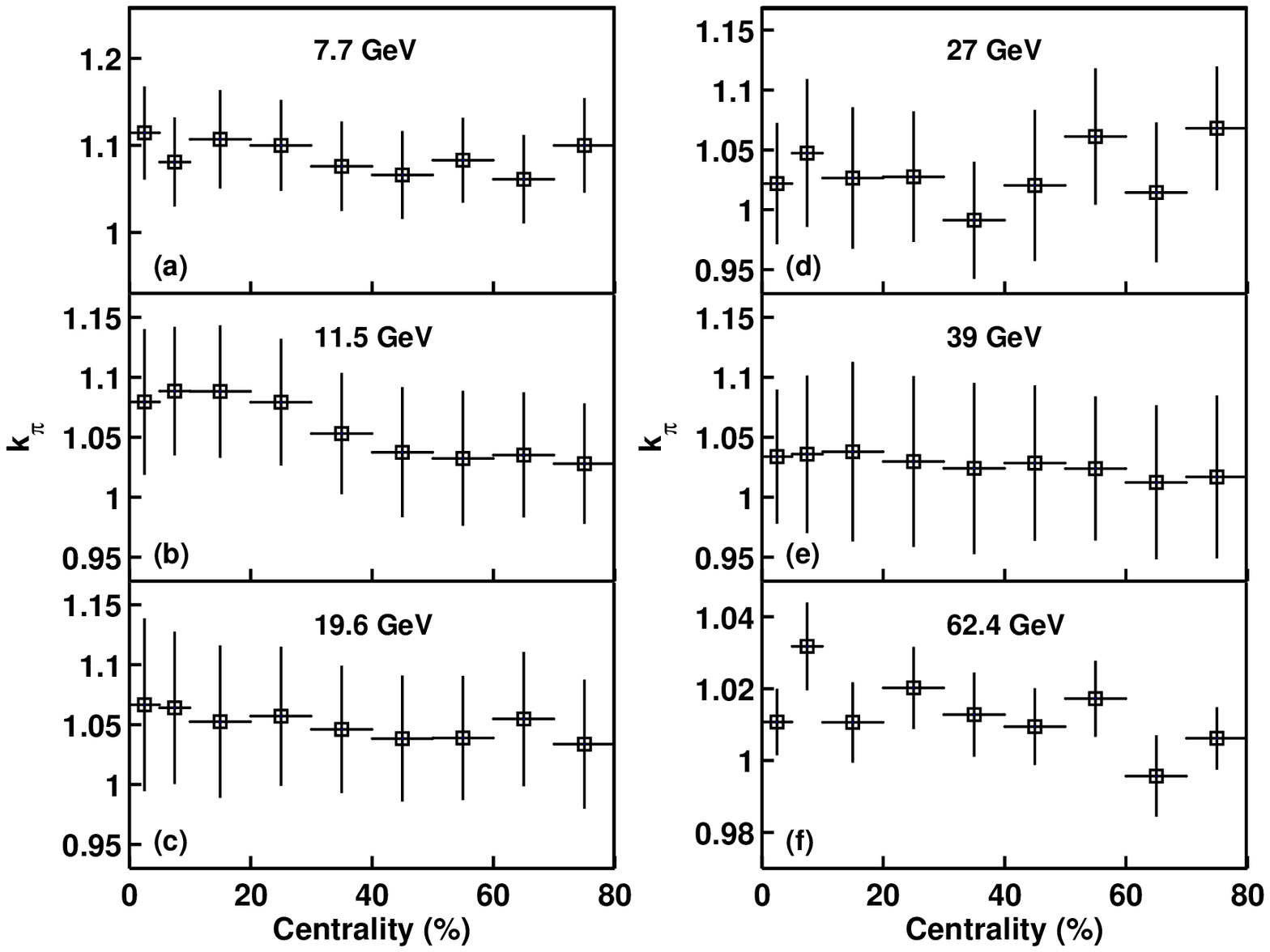}}
\vskip-0.18cm  Figure 7. Centrality-dependent $k_{\pi}$ at different energies of (a) 7.7, (b) 11.5, (c) 19.6, (d) 27, (e) 39 and (f) 62.4 GeV.  
\end{figure}

\begin{figure}[H]
\hskip-0.0cm {\centering
\includegraphics[width=16.0cm]{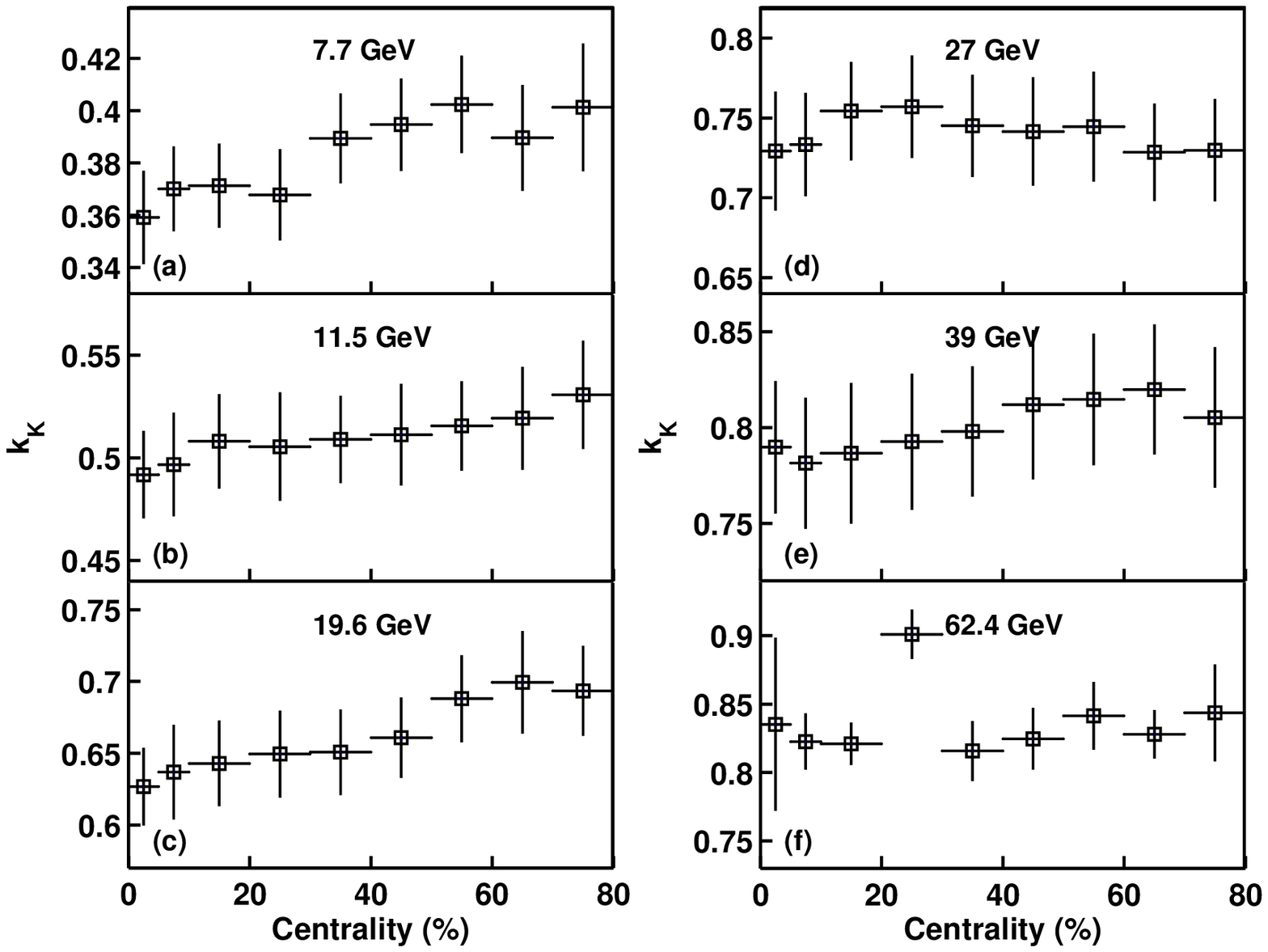}}
\vskip-0.18cm  Figure 8. Same as Figure 7 but for $k_{K}$.  
\end{figure}

\begin{figure}[H]
\hskip-0.0cm {\centering
\includegraphics[width=16.0cm]{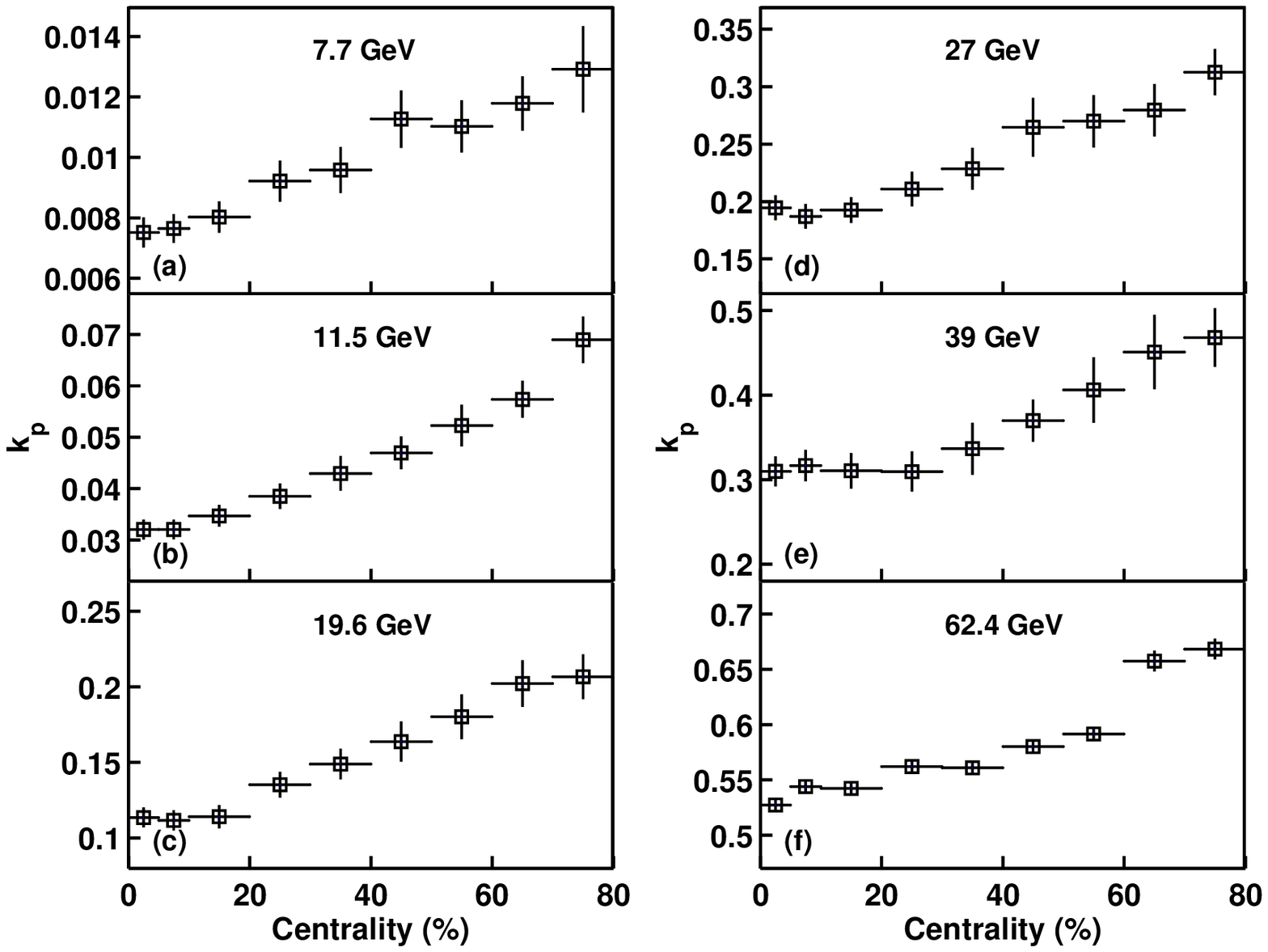}}
\vskip-0.18cm  Figure 9. Same as Figure 7 but for $k_{p}$.  
\end{figure}
 
In Figures 7--9, one can see that the three types of yield ratios obviously depend on collision energy, and we find that the logarithms of the three types of yield ratios, $\ln(k_{\pi})$, $\ln(k_{K})$, and $\ln(k_{p})$, show obvious linear dependence on $1/\sqrt{s_{NN}}$, and the linear relationship can be expressed as
\begin{gather}
\ln(k_{ij})=A_{ij}/\sqrt{s_{NN}}+B_{ij},
\end{gather}
where $i$ represents $\pi$, $K$, or $p$, $j$ represents different centrality classes, and $A_{ij}$ and $B_{ij}$ are fitting parameters.   
Figure 10 shows the $1/\sqrt{s_{NN}}$-dependent (a) $\ln(k_{\pi})$, (b) $\ln(k_{K})$, and (c) $\ln(k_{p})$ for different centralities. The fitting lines are the results calculated by Least Squares Method.
The values of calculated parameters ($A_{ij}$ and $B_{ij}$) and $\chi^2$/dof are listed in Table 7. It is not hard to see that, the values of intercept $B_{ij}$ are asymptotically zero, which means the limiting values of the yield ratios are one at very high energy. For the same particle, the slope $A_{ij}$ does not change much with the increase of centrality, especially for $\pi$. To see clearly the dependences of the linear relationships on centrality, the results for different centrality classes are added or subtracted by appropriate factors shown in different panels of Figure 10.  
\begin{figure}[H]
\hskip-0.0cm {\centering
\includegraphics[width=16.0cm]{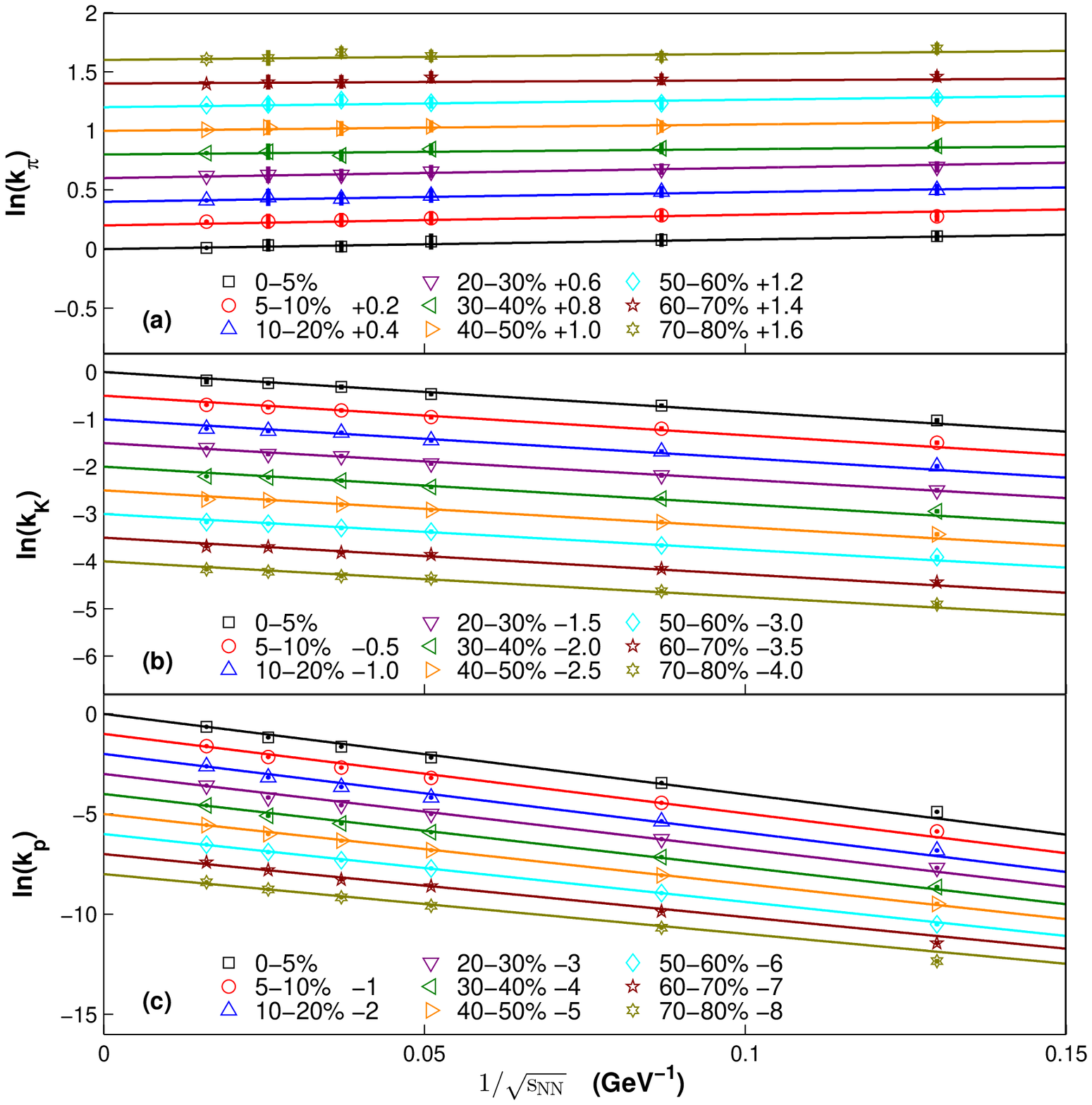}}
\vskip-0.18cm  Figure 10. Energy dependent (a) $\ln(k_{\pi})$, (b) $\ln(k_{K})$, and (c) $\ln(k_{p})$ for different centralities. The fitting lines are the results calculated by Least Squares Method. For clarity, the results for different centralities are added or subtracted by appropriate factors shown in different panels.
\end{figure}

\newpage
{\scriptsize {Table 7. Values of free parameters and $\chi^2$/dof  corresponding to the fitting lines in Figure 10.
{%
\begin{center}
\begin{tabular}{cccccc}
\hline
\hline
 Figure& Particle & Centrality &  $A_{ij}$ & $B_{ij}$ & $\chi^2$/dof \\
\hline
&& 0-5\% &    0.886 $\pm$    0.399 &   -0.003 $\pm$    0.010 &    0.209/3\\
&& 5-10\% &    0.507 $\pm$    0.376 &    0.024 $\pm$    0.013 &    0.229/3\\
&& 10-20\% &    0.880 $\pm$    0.459 &   -0.003 $\pm$    0.012 &    0.171/3\\
&& 20-30\% &    0.703 $\pm$    0.403 &    0.009 $\pm$    0.012 &    0.086/3\\
Figure 10(a)&  $\pi$ & 30-40\% &    0.544 $\pm$    0.370 &    0.003 $\pm$    0.013 &    0.516/3\\
&& 40-50\% &    0.482 $\pm$    0.370 &    0.002 $\pm$    0.011 &    0.114/3\\
&& 50-60\% &    0.500 $\pm$    0.379 &    0.010 $\pm$    0.012 &    0.511/3\\
&& 60-70\% &    0.613 $\pm$    0.426 &   -0.013 $\pm$    0.013 &    0.507/3\\
&& 70-80\% &    0.692 $\pm$    0.331 &   -0.004 $\pm$    0.010 &    1.231/3\\
\hline
&& 0-5\% &   -7.522 $\pm$    0.444 &   -0.055 $\pm$    0.032 &    0.656/3\\
&& 5-10\% &   -7.086 $\pm$    0.349 &   -0.075 $\pm$    0.028 &    0.622/3\\
&& 10-20\% &   -6.956 $\pm$    0.424 &   -0.076 $\pm$    0.030 &    2.076/3\\
&& 20-30\% &   -7.922 $\pm$    0.501 &    0.011 $\pm$    0.036 &    1.985/3\\
Figure 10(b)&{K}& 30-40\% &   -6.667 $\pm$    0.313 &   -0.081 $\pm$    0.027 &    1.503/3\\
&& 40-50\% &   -6.641 $\pm$    0.324 &   -0.073 $\pm$    0.026 &    1.162/3\\
&& 50-60\% &   -6.688 $\pm$    0.323 &   -0.053 $\pm$    0.023 &    1.122/3\\
&& 60-70\% &   -6.637 $\pm$    0.499 &   -0.068 $\pm$    0.040 &    2.414/3\\
&& 70-80\% &   -6.570 $\pm$    0.496 &   -0.056 $\pm$    0.029 &    0.534/3\\
\hline
&& 0-5\% &  -38.697 $\pm$    1.583 &   -0.031 $\pm$    0.138 &   32.951/3\\
&& 5-10\% &  -38.855 $\pm$    3.000 &   -0.002 $\pm$    0.200 &   46.465/3\\
&& 10-20\% &  -38.147 $\pm$    1.051 &   -0.014 $\pm$    0.092 &   37.401/3\\
&& 20-30\% &  -37.168 $\pm$    0.600 &    0.010 $\pm$    0.029 &   23.505/3\\
Figure 10(c)& $p$ & 30-40\% &  -36.214 $\pm$    0.725 &   -0.001 $\pm$    0.012 &    7.260/3\\
&& 40-50\% &  -35.067 $\pm$    0.523 &    0.014 $\pm$    0.014 &    3.675/3\\
&& 50-60\% &  -34.785 $\pm$    0.479 &    0.032 $\pm$    0.010 &    1.071/3\\
&& 60-70\% &  -35.004 $\pm$    0.530 &    0.140 $\pm$    0.016 &    3.368/3\\
&& 70-80\% &  -33.417 $\pm$    0.978 &    0.131 $\pm$    0.016 &    4.916/3\\
\hline
\hline
\end{tabular}
\end{center}
}} }

As can be seen, with increase of $\sqrt{s_{NN}}$, $\ln(k_{K})$ and $\ln(k_{p})$ increase, while $\ln(k_{\pi})$ decreases, which implies that the generation mechanism of $K$ is similar to $p$, but is different from  $\pi$. The differences of cross-section of absorbtion, content of primary proton in nuclei and so on can result in the differences of the yields of these particles. From Table 7, one can see that the centrality-dependent $A_{\pi}$ varies roughly between 0.48 and 0.89, and does not show obvious change trend with centrality. $A_{K}$ and $A_{p}$ vary roughly from -7.52 to -6.63 and from -38.86 to -33.42, respectively, and overall decrease with centrality, and the increase of $A_{p}$ is relatively obvious. These indicate that although the dependence of the energy-dependent yield ratio of $p$ on centrality is higher than that of $K$, and that of $K$ is higher than that of $\pi$, the dependences of the three energy-dependent yield ratios on centrality are not obvious.

Based on the extracted yield ratios and Equations (9) and (10), the energy- and centrality-dependent light hadron chemical potentials, $\mu_{\pi}$, $\mu_{K}$, and $\mu_{p}$, of $\pi$, $K$, and $p$, and quark chemical potentials, $\mu_{u}$, $\mu_{d}$, and $\mu_{s}$, of $u$, $d$, and $s$ quarks, are obtained and shown in Figure 11. The different symbols denote the calculated results of different centrality classes. The curves are the derivative results according to Equation (11) corresponding to the fitted lines in Figure 10. As can be seen, in the energy range from 7.7 to 62.4 GeV, $\mu_{\pi}$ increases, and $\mu_{K}$, $\mu_{p}$, $\mu_{u}$, $\mu_{d}$, and $\mu_{s}$  decrease obviously with the increase of $\sqrt{s_{NN}}$. From the trends of curves, the limiting values of the six types of chemical potentials are asymptotically zero at very high energy. The differences between chemical potentials of particles with different centralities are relatively large in low energy region, and as the energy increases, the differences gradually decrease, and finally tend to be zero at very high energy. Overall, the energy-dependent chemical potentials of hadrons and quarks in different centrality classes are very close, which means that the six types of energy-dependent chemical potentials from Au-Au collisions have a little dependences on centrality. In addition, at the same energy, $\mu_{K}$ is larger than $|\mu_{\pi}|$, but less than $\mu_{p}$, and $\mu_{u}$ is almost as large as $\mu_{d}$, but larger than $\mu_{s}$ due to the differences of different particle masses.

It is not hard to notice that $\mu_{\pi} < 0$, while $\mu_{K}(\mu_{p},\mu_{u}, \mu_{d}, \mu_{s}) > 0$. This is caused by $k_{\pi} > 1$, while $k_{K}(k_{p}) < 1$. When the energy increases to a very high value, all
chemical potentials of light hadrons and quarks approach to zero,  when the partonic interactions possibly play a dominant role, mean-free-path of particles becomes large, and the collision system possibly changes completely from the hadron-dominant state to the quark-dominant state.
\begin{figure}[H]
\hskip-0.0cm {\centering
\includegraphics[width=16.0cm]{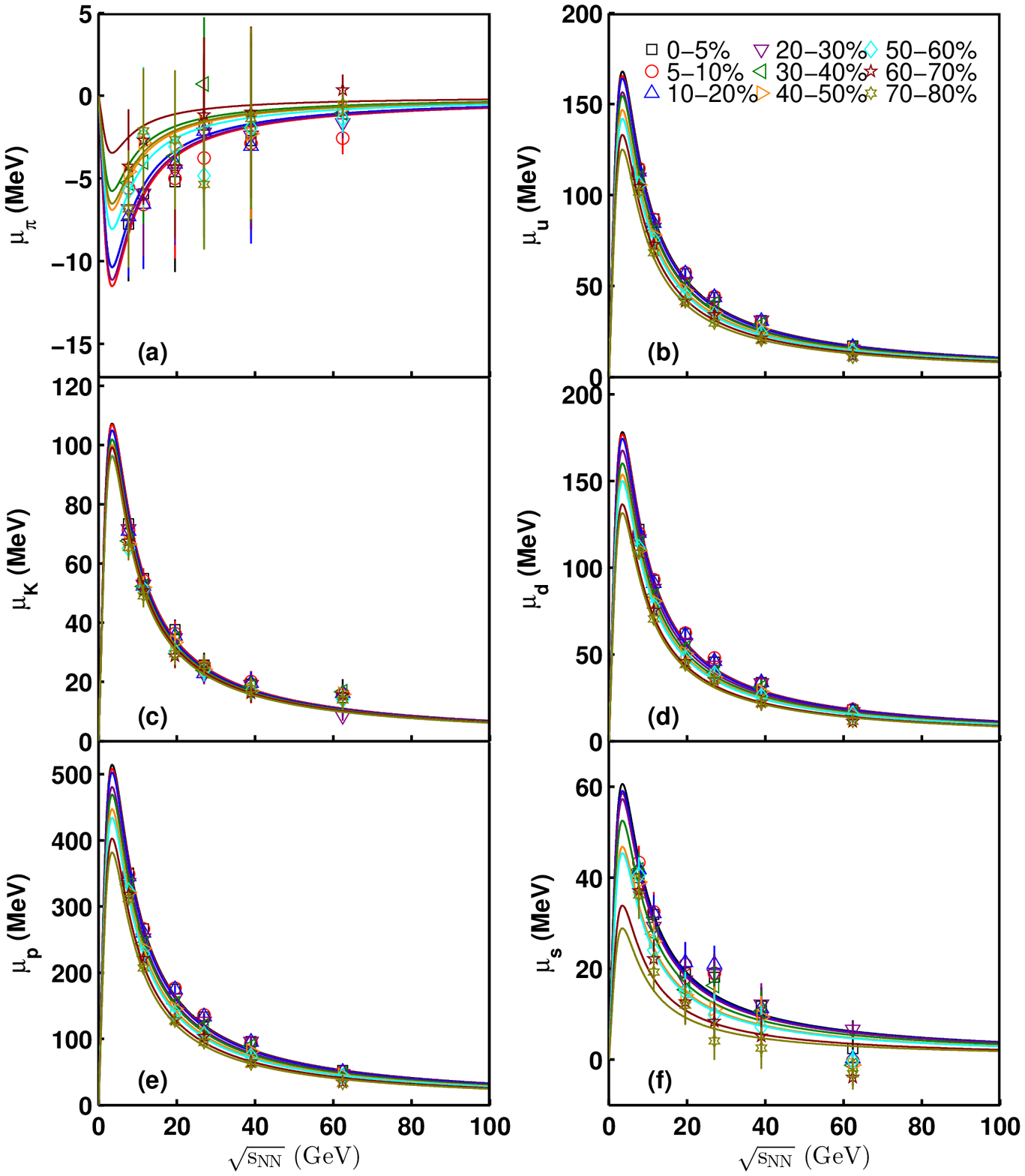}}
\vskip-0.18cm  Figure 11. Dependences of centrality-dependent hadron chemical potentials, (a) $\mu_{\pi}$, (b) $\mu_{K}$, and (c) $\mu_{p}$ and quark chemical potentials, (d) $\mu_{u}$, (e)$\mu_{d}$, and (f) $\mu_{s}$, on energy $\sqrt{s_{NN}}$. The different symbols denote the calculated results in different centrality classes according to the extracted yield ratios and Equations (9) and (10), and the curves are the derivative results based on the linear fits in Figure 10.
\end{figure}
In Figure 11, like our previous work, the derived curves of  particle chemical potentials from the linear fits of the energy-dependent yield ratios in Figure 10 simultaneously show the maximum (the absolute magnitude for $\pi$ ) at 3.526 GeV, which is not observed from the linear fits. The energy at the maximum can be obtained according to the calculation method in reference. According to Equations (7), and (9--11), we can obtain the chemical potentials ($\mu_{kj}$) of hadrons and quarks for all centrality classes in terms of $\sqrt{s_{NN}}$. Considering that all the values of intercept ${B_{ij}}$ in Table 7 approximate to zero and the simplicity for calculation, we set ${B_{ij}}=0$ here, and the 
$\mu_{kj}$ can be written as
\begin{equation}
\mu_{kj}=T_{ch}\frac{C_{kj}}{\sqrt{s_{NN}}}, \\
\end{equation}
Where $k=\pi$, $K$, $p$, $u$, $d$, and $s$, $j$ represents different centralities, and $C_{kj}$ is a linear combination of $A_{ij}$, i.e. 
\begin{gather}
\left\{
\begin{aligned}
&C_{{\pi}j}=-\frac{1}{2}A_{{\pi}j},\hfill\\
&C_{{K}j}=-\frac{1}{2}A_{{K}j},\hfill\\
&C_{{p}j}=-\frac{1}{2}A_{{p}j},\hfill\\
&C_{{u}j}=-\frac{1}{6}(A_{{\pi}j}+A_{{p}j}),\hfill\\
&C_{{d}j}=-\frac{1}{6}(-2A_{{\pi}j}+A_{{p}j}),\hfill\\
&C_{{s}j}=-\frac{1}{6}(A_{{\pi}j}-2A_{{K}j}+A_{{p}j}).\hfill
\end{aligned}
\right.
\end{gather}
Let $\frac{\mathrm{d}\mu_{ij}}{\mathrm{d}\sqrt{s_{NN}}}=0$, we obtain the energy value ($\sqrt{s_{NN}}=3.526$ GeV) at the maximum.

It must be emphasized that, due to the lack of data in low-energy region, the maximum here is only a prediction according to these linear fits, not a certainty. The energy (3.526 GeV) at the maximum possibly is the critical energy of phase transition from a liquid-like hadron state to a gas-like quark state in the collision system. At this special energy, the chemical potentials for all cases have the maximum, which indicates that the density of baryon number has the largest value and the mean-free-path of particles has the smallest value. That means the hadronic interactions play an important role at this special stage. When the energy is higher than 3.526 GeV, these chemical potentials gradually decrease with the increase of energy, which indicates that the density of baryon number gradually decreases, the mean-free-path increases, the shear viscosity over entropy density gradually weakens, the hadronic interactions gradually fade, and the partonic interactions gradually become greater. When the energy increases to a very high value, especially the LHC energy, the chemical potentials of all types of particles approach to zero, which means that the density of baryon number  and the viscous effect approach to zero, and the collision system possibly changes completely from the hadron-dominant state to the quark-dominant state, which denotes the partonic interactions possibly play a dominant role at very high energy, and the strongly coupled QGP (sQGP) has been observed.

we must point out that, since the maximum is predicted by the empirical formula of fit, there is a large fluctuation in the extracted critical energy value. In other words, although the fluctuation exists or is even large, it does not mean that the extracted energy value must be wrong. So at the extracted energy, there may be a phase transition critical point. The extracted critical energy (3.526 GeV) of phase transition is consistent with our previous result and the result (below 19.6 GeV) by the STAR Collaboration, but less than the result (between 11.5 GeV and 19.6 GeV) of a study based on a correlation between collision energy and transverse momentum. It can be seen that we still need to make more efforts to find the critical energy point through new methods or theories.

{\section{Summary and conclusion}}

The transverse momentum spectra of final-state light flavour particles, $\pi^{\pm}$, $K^{\pm}$, $p$, and $\bar p$, produced in Au-Au collisions for different centralities over an energy range from 7.7 to 62.4 GeV, are described by a two-component Erlang distribution in the frame of multi-source thermal model. The fitting results are in agreement with the experimental data recorded by the STAR Collaboration.

From the fitting parameters of two-component Erlang $p_{T}$ distribution, the first component corresponding to a narrow low-$p_{T}$ region, is contributed by the soft excitation process where a few (2--4) sea quarks and gluons take part in, and the second component corresponding to a wide high-$p_{T}$ region, is contributed by the hard scattering process coming from a more violent collision between two valent quarks in incident nucleons. The relative weight factor of soft excitation process shows that the contribution ratio of soft excitation process is more than 60\%, which indicates that the excitation degree of collision system is mainly contributed by the soft excitation process.

The yield ratios, $k_{\pi}$, $k_{K}$, and $k_{p}$, of negative to positive particle versus collision energy and centrality are obtained from the normalization constants. The study shows that, although the dependence of $k_{p}$ on centrality is higher than that of $k_{K}$, and the dependence of $k_{K}$ on centrality is higher than that of $k_{\pi}$, the dependences of the three yield ratios on centrality are not significant, especially for $\pi$. The logarithms of the three types of yield ratios, $\ln(k_{\pi})$, $\ln(k_{K})$, and $\ln(k_{p})$, show obvious linear dependence on $1/\sqrt{s_{NN}}$. 

The energy- and centrality-dependent chemical potentials of light hadrons, $\mu_{\pi}$, $\mu_{K}$, and $\mu_{p}$, and quarks, $\mu_{u}$, $\mu_{d}$, and $\mu_{s}$, are extracted from the yield ratios. With the increase of energy over a range from 7.7 to 62.4 GeV, all the chemical potentials (the absolute magnitude for $\pi$) decrease obviously. When the collision energy is very high, all types of chemical potentials are small and tend to be a limiting value of zero. Overall, the dependences of the six types of energy-dependent chemical potentials on centrality are relatively more obvious in low energy region than that in high energy region, but the six energy-dependent chemical potentials in different centrality classes are very close, which indicates that the dependences of the energy-dependent chemical potentials from Au-Au collisions on centrality are relatively not so significant. 

All the derived curves of energy- and centrality-dependent chemical potentials of hadrons and quarks, based on the linear relationships between the logarithms of yield ratios and $1/\sqrt{s_{NN}}$, simultaneously show the maximum (the absolute magnitude for $\pi$) at 3.526 GeV, which possibly is the critical energy of phase transition from a liquid-like hadron state to a gas-like quark state in the collision system, when the density of baryon number in Au-Au collisions has a large value and the hadronic interactions play an important role. When energy continues to increase, all types of chemical potentials become small, which indicates the density of baryon number gradually decreases, the mean-free-path gradually increases, and the viscous effect gradually weakens, when the hadronic interactions gradually fade and the partonic interactions gradually become greater. When the energy rises to a very high value, especially to the LHC, all types of chemical potentials tend to zero, which indicates that the collision system possibly changes completely from the liquid-like hadron-dominant state to the gas-like quark-dominant state, when the partonic interactions possibly play a dominant role. \\

{\bf Acknowledgments}

This work was supported by the National Natural Science Foundation of China under Grant Nos. 11847114, 11575103 and 11775104, the Shanxi Provincial Natural Science Foundation under Grant No.
201701D121005, and the Fund for Shanxi ``1331 Project" Key Subjects Construction.\\

{\bf Data availability}

All data are quoted from the mentioned references. As a phenomenological work, this paper does not report new data.\\

{\bf Conflicts of Interest}

The authors declare that there are no conflicts of interest regarding the publication of this paper.\\

\end{document}